\documentclass[prd,superscriptaddress,amsfonts,amssymb,amsmath,showpacs,onecolumn]{revtex4-2}

\usepackage{bm}
\usepackage{amsfonts}
\usepackage{latexsym}
\usepackage{graphicx}
\usepackage{amsmath}
\usepackage{palatino}
\usepackage{mathpazo}
\usepackage{textcomp}
\usepackage{float}
\usepackage{dcolumn}
\usepackage{booktabs}
\usepackage{multirow}
\usepackage{hyperref}
\usepackage{orcidlink}
\hypersetup{colorlinks,citecolor=blue}
\usepackage[caption=false]{subfig}
\usepackage{xcolor}

\begin{document}
	\raggedbottom
	\color{black}       
	\title{Running Vacuum Cosmology in f(R,T) Gravity: Observational Constraints and Thermodynamic Analysis}
	
	\author{Anjani\orcidlink{0009-0000-1108-7648}}
	\email{anjaliguleria002@gmail.com}
	\affiliation{Department of Mathematics, Maharaja Agrasen University, Kalujhanda, Solan, Himachal Pradesh-174103, India} 
	
	\author{Ajay Kumar\orcidlink{0000-0002-9478-2175}}
	\email{maths.ajy@gmail.com}
	\affiliation{School of Technology Management \& Engineering,
		SVKM's NMIMS (Deemed to be University),
		Chandigarh-160014, India.} 
		
	\author{Pinki\orcidlink{0009-0000-8606-0423}}
	\email{shirdpinki@gmail.com}
	\affiliation{Department of Mathematics, GMN College, Ambala Cantt.-133001, Haryana, India.}
	
	\author{Pankaj Kumar\orcidlink{0000-0002-7156-5637}}
	\email{pankaj.11dtu@gmail.com}
	\affiliation{Department of Mathematics, Maharaja Agrasen University, Kalujhanda, Solan, Himachal Pradesh-174103, India} 
	\begin{abstract}	
Despite the remarkable observational success of the $\Lambda$CDM model, the physical origin of dark energy and the cosmological constant remain unresolved, motivating the exploration of dynamical vacuum scenarios within modified theories of gravity. Inspired by the quantum field theoretical description of running vacuum energy, we investigate a running vacuum cosmology in the framework of linear $f(R,T)=R+\lambda T$ gravity. Unlike recent studies that establish an effective correspondence between running vacuum energy and modified gravity, our approach directly incorporates a quantum field theory motivated running vacuum scenario into the $f(R,T)$ framework and examines its cosmological implications. Exact analytical solutions of the modified field equations are obtained in a spatially flat Friedmann--Lemaître--Robertson--Walker spacetime, yielding an analytical expression for the Hubble parameter. The free model parameters are constrained using the latest Pantheon+SH0ES compilation, DESI baryon acoustic oscillation measurements, and their joint dataset through Markov Chain Monte Carlo analysis. The observationally constrained model successfully reproduces the observed expansion history, remains consistent with independent Cosmic Chronometer measurements, predicts an age of the universe within current observational bounds, and yields an effective equation of state compatible with current observations. Furthermore, the statefinder and $Om$ diagnostics indicate close agreement with the $\Lambda$CDM scenario at late times, while the generalized second law of thermodynamics remains valid throughout cosmic evolution. These results demonstrate that the proposed framework provides an observationally viable and thermodynamically consistent realization of running vacuum cosmology in $f(R,T)$ gravity, offering a theoretically motivated alternative for describing the present accelerated universe.
\end{abstract}
	\maketitle
	\newpage
\section{Introduction}
The current understanding of our universe is founded on two remarkable discoveries: the geometric description of gravity and the expansion of the cosmos. The former is successfully explained by General Relativity (GR), which has provided an accurate description of a wide range of gravitational phenomena, from planetary motion and black holes to the recent direct detection of gravitational waves \cite{Will,Abbott}. The latter was first established by Edwin Hubble through astronomical observations, while the discovery of the late-time accelerated expansion of the universe at the end of the twentieth century revolutionized modern cosmology. Independent observations of distant Type Ia supernovae (SNe Ia) demonstrated that the cosmic expansion is accelerating rather than decelerating under the influence of gravity \cite{1,Per1}. This remarkable discovery has subsequently been confirmed by several independent cosmological probes, including Baryon Acoustic Oscillations (BAO), large-scale structure surveys, and high-precision measurements of the Cosmic Microwave Background (CMB) from the Wilkinson Microwave Anisotropy Probe (WMAP) and the Planck mission \cite{Spergel1,Tegmark1,Eienstein,Planck}. According to the latest cosmological observations, nearly $68.3\%$ of the total energy density of the universe is attributed to dark energy (DE), about $26.8\%$ consists of dark matter (DM), while ordinary baryonic matter contributes only approximately $4.9\%$.

The discovery of cosmic acceleration immediately raised one of the most fundamental questions in modern cosmology: what is the physical origin of the present accelerated expansion of the universe? This observational evidence suggests either the existence of an exotic cosmic component possessing sufficiently negative pressure, collectively known as DE, or a possible breakdown of GR on cosmological scales requiring modifications of the gravitational theory itself. Within the framework of GR, the simplest explanation is provided by the cosmological constant ($\Lambda$), which, together with Cold Dark Matter (CDM), constitutes the standard $\Lambda$CDM cosmological model. Owing to its remarkable agreement with a wide variety of cosmological observations, the $\Lambda$CDM model has become the reference framework for describing the large-scale evolution of the universe \cite{Alam}. Nevertheless, despite its observational success, the model continues to face several long-standing theoretical challenges, including the unknown physical origin of DE, the cosmological constant problem, and the cosmic coincidence problem \cite{coscon,coin}.

Motivated by these shortcomings, numerous dynamical DE models have been proposed in which the equation of state evolves with cosmic time or redshift. Among the most extensively studied candidates are quintessence \cite{quin1,quin2}, k-essence \cite{kess1,kess2}, tachyonic fields \cite{tac}, and phantom models \cite{phan}. Although these scalar-field models successfully reproduce the observed late-time accelerated expansion, they generally suffer from several theoretical limitations, including fine-tuning, sensitivity to initial conditions, the absence of a compelling fundamental origin for the scalar fields, and possible stability issues at the perturbative level \cite{prob}. Moreover, the large degeneracy among different scalar-field models makes them difficult to distinguish using current observational data \cite{prob1}. These limitations have motivated an alternative perspective in which the geometric sector of gravity, rather than the matter sector, is modified. Such modifications naturally give rise to a broad class of modified gravity theories capable of explaining the observed cosmic acceleration without introducing additional exotic DE components.

Over the past few decades, a variety of modified theories of gravity have been proposed as possible extensions of Einstein's GR. A natural generalization is provided by $f(R)$ gravity, in which the Einstein--Hilbert action is extended by replacing the Ricci scalar $R$ with an arbitrary function $f(R)$ formulated within the standard Riemannian geometry \cite{fr,fr1}. This idea has subsequently been generalized by incorporating higher-order curvature invariants through $f(R,G)$ gravity, where $G$ denotes the Gauss--Bonnet invariant, leading to richer geometric corrections to the gravitational action \cite{frg,frg1}. Besides curvature-based formulations, gravity can also be described through alternative geometric quantities such as torsion and non-metricity. This has resulted in the development of teleparallel $f(T)$ gravity, its extension $f(T,B)$ (where $B$ is the boundary term), and the symmetric teleparallel formulation of $f(Q)$ gravity, all of which have been extensively investigated as viable alternatives to GR \cite{ft,ftb,fq}.

An important class of modified gravity theories involves an explicit coupling between matter and geometry. Representative examples include $f(R,T)$ and $f(Q,T)$ gravity, where $T$ denotes the trace of the energy--momentum tensor \cite{Harko,fqt}. Owing to the matter--geometry coupling, these theories naturally modify the cosmic fluid dynamics and may generate an effective negative pressure capable of driving the late-time accelerated expansion without introducing an additional DE component. Among these frameworks, the $f(R,T)$ theory proposed by Harko \textit{et al.} \cite{Harko} has attracted considerable attention because the gravitational action depends explicitly on both the Ricci scalar $R$ and the trace of the energy--momentum tensor $T$. Such a coupling may originate from quantum effects or imperfect matter sources and provides a simple yet rich extension of GR. Consequently, $f(R,T)$ gravity has been widely employed to investigate different cosmological scenarios. Houndjo reconstructed suitable $f(R,T)$ models capable of reproducing different evolutionary phases of the universe \cite{Houndjo}. The physical viability of these models has further been examined through stability analyses and energy conditions \cite{Shariff,estab}. The effects of bulk viscosity and particle creation within the $f(R,T)$ framework have also been extensively explored \cite{t2,cps}. Moraes and Santos developed the $f(R,T^\phi)$ formalism by incorporating scalar fields into the matter--geometry coupling \cite{Moraes}, while Sharif and Siddiqa investigated the thermodynamic behaviour of an FLRW universe in $f(R,T)$ gravity \cite{Sharif}. Several other investigations have demonstrated that $f(R,T)$ gravity successfully describes the transition from an early decelerated phase to the present accelerated epoch in agreement with observational data \cite{trans,tran1,trans3}. Furthermore, the theory has also been applied successfully to non-singular wormhole geometries, where the trace-dependent terms play an essential role in sustaining traversable wormhole solutions \cite{worm}.

The standard $\Lambda$CDM model assumes that the cosmological constant remains strictly constant throughout the entire cosmic evolution. However, quantum field theory predicts a vacuum energy density exceeding the observed value by nearly $10^{120}$ orders of magnitude, giving rise to the well-known cosmological constant problem. In addition, the cosmic coincidence problem and the persistent discrepancy between early and late universe measurements of the Hubble constant further motivate the exploration of cosmological models with a dynamical vacuum sector \cite{hubten,allp}. Among the various possibilities, models with a time-dependent cosmological constant have received considerable attention because they provide a natural framework for describing vacuum evolution while potentially alleviating some of these long-standing theoretical and observational difficulties.

The idea of a decaying vacuum is not new and has evolved considerably over the past several decades. One of the earliest phenomenological models was proposed by Ozer and Taha \cite{Ozer} as a possible solution to several shortcomings of the standard cosmological model. Shortly afterwards, Freese \textit{et al.} demonstrated that a decaying vacuum could naturally act as a source of particle production during cosmic evolution \cite{Freese}. Chen and Wu \cite{Chen} subsequently argued that the vacuum energy may originate from quantum cosmological considerations and proposed the simple decay law $\Lambda\propto H^2$. This framework was further generalized by Carvalho \textit{et al.} \cite{Carvalho}, who introduced a more general phenomenological vacuum model capable of describing the transition from the inflationary era to the radiation-dominated epoch. A major theoretical advance came with the work of Shapiro and Sol\`a \cite{Shapiro}, who demonstrated that the vacuum energy density is expected to evolve with the expansion of the universe within the framework of quantum field theory in curved spacetime through the renormalization group approach. In this picture, the vacuum energy is no longer regarded as a strict constant but rather as a slowly evolving quantity whose dominant late-time contribution is proportional to the square of the Hubble parameter. Subsequent theoretical and observational investigations have shown that running vacuum cosmology provides an excellent description of the late-time expansion history and remains highly consistent with current cosmological observations \cite{Gomez,obs}.

Recently, Errahmani \textit{et al.} \cite{Errahmani} investigated the cosmological constant within the framework of $f(R,T)$ gravity by considering the gravitational action $f(R,T)=R+2\lambda\kappa^2T-2\Lambda$. They demonstrated that, owing to the intrinsic matter--geometry coupling, the resulting cosmological dynamics can effectively reproduce the behaviour of a running vacuum model and established an interesting correspondence between this particular class of $f(R,T)$ gravity and running vacuum cosmology. The objective of the present work, however, is fundamentally different. Rather than deriving an effective running vacuum behaviour from the modified gravitational action, we directly investigate a quantum field theory motivated running vacuum model within the framework of $f(R,T)=R+\lambda T$ gravity. In this approach, the running vacuum constitutes the physical DE sector, while the matter--geometry coupling modifies the underlying gravitational dynamics. This allows us to investigate the combined cosmological effects of a quantum-inspired dynamical vacuum and modified gravity within a unified framework. Throughout the present work, we restrict our attention to the late-time universe and adopt the commonly used phenomenological form of the running vacuum, in which the dominant contribution to the vacuum energy density is proportional to $H^2$ \cite{jsp}.

To examine the observational viability of the proposed model, we constrain the free parameters using the latest Pantheon+SH0ES SNe Ia compilation, the DESI 2024 BAO measurements, and their joint dataset. The parameter estimation is performed using the Markov Chain Monte Carlo (MCMC) technique. We investigate the cosmological evolution of the model through the Hubble parameter, deceleration parameter, effective equation of state, age of the universe, statefinder, and $Om$ diagnostics. In addition, the observational consistency of the proposed running vacuum model is assessed through a residual analysis using Cosmic Chronometer data, while its thermodynamic viability is examined via the generalized second law of thermodynamics. These analyses provide a comprehensive assessment of the observational consistency and physical viability of the proposed running vacuum cosmology in $f(R,T)$ gravity.

The paper is organized as follows. In Section~2, we present the field equations of $f(R,T)$ gravity in a spatially flat Friedmann--Lema\^{\i}tre--Robertson--Walker universe and derive the analytical solution for the Hubble parameter. Section~3 briefly describes the observational datasets employed in the analysis. The observational constraints and the corresponding best-fit values of the model parameters are presented in Section~4. In Section~5, we investigate the cosmological evolution of the Hubble parameter, deceleration parameter, effective equation of state, age of the universe, and the residual analysis. Section~6 is devoted to the statefinder and $Om$ diagnostic analyses. The thermodynamic behaviour of the model is discussed in Section~7. Finally, the main conclusions of the present work are summarized in Section~8.
 
\section{$f(R,T)$ Theory and Its Field Equations}
The Einstein--Hilbert action for $f(R,T)$ gravity, adopting units in which $8\pi G=c=1$, is given by
\begin{equation}
S=\frac{1}{2}\int[f(R,T)+2\mathcal{L}_{m}]\sqrt{-g} d^4x ,
\end{equation}
where $\mathcal{L}_{m}$ is the matter Lagrangian density, $g$ is the determinant of the metric tensor $g_{\mu\nu}$, and $R$ denotes the Ricci scalar. Furthermore, $T=g^{\mu\nu}T_{\mu\nu}$ represents the trace of the matter energy--momentum tensor and is defined as
\begin{equation}
T_{\mu\nu}= \frac{-2}{\sqrt{-g}} \frac{\delta ( \sqrt{-g} \mathcal {L}_{m})}{\delta g^{\mu\nu}}
\end{equation}
In $f(R,T)$ gravity, the Lagrangian density of matter $\mathcal L_{m}$ depends only on  $g_{\mu\nu}$ and not on its derivatives. Therefore, Eq.~(2) becomes
\begin{equation}
T_{\mu\nu}= g_{\mu\nu}\mathcal{L}_{m} - 2\frac{\partial\mathcal L_{m}}{\partial  g^{\mu\nu}}
\end{equation}

An interesting feature of this theory is the non-conservation of the energy--momentum tensor, i.e., $\nabla^{\mu}T_{\mu\nu}\neq0$, which indicates a possible exchange of energy between matter and geometry. From a thermodynamic perspective, this non-conservation can be interpreted as an effective mechanism for particle production in an open cosmological system \cite{Harko14}. The variation of the gravitational action $S$ given in Eq.~(1) with respect to the metric tensor $g^{\mu\nu}$ yields the modified field equation in metric form as
\begin{equation}
f_{R}(R,T) R_{\mu\nu}  -  \frac{1}{2} f(R,T) g_{\mu\nu} +  ( g_{\mu\nu} \Box  -  \nabla_{\mu} \nabla_{\nu}) f_{R}(R,T)  = T_{\mu\nu} -  f_{T}(R,T) T_{\mu\nu}  -  f_{T}(R,T) \Theta_{\mu\nu},
\end{equation}
where $f_{R}(R, T )=\frac{ \partial f(R,T)}{\partial R}$ and $ f_{T}(R,T)=\frac{\partial  f(R,T)}{\partial T}$, $\nabla_{\mu}$ denotes the covariant derivative w.r.t $g_{\mu\nu}$ and $\Box$ is the D'Alembert operator. The term $ \Theta_{\mu\nu} $ appearing in the field equation describes the dependence of matter on metric tensor and can be written as
	\begin{equation}
		\Theta_{\mu\nu} = -2T_{\mu\nu} + g_{\mu\nu}\mathcal L_{m} - 2g^{\alpha\beta} \frac{\partial^{2}\mathcal L_{m}}{\partial g^{\mu\nu}\partial g^{\alpha\beta}}.
	\end{equation}
	
In general, the gravitational field in $f(R,T)$ is influenced by the physical nature of the matter source through the tensor $\Theta_{\mu\nu}$. The functional freedom with arbitrary function in $f(R,T)$ gravity allows the construction of several viable and different matter models in theoretical framework. The well-known functional forms include
\begin{equation*}
	f(R,T) =
	\begin{cases}
		R + f(T) \\
		f_{1}(R) + f_{2}(T) \\
		f_{1}(R) + f_{2}(R)f_{3}(T)
	\end{cases}
\end{equation*}
such that $f_{n}(R)$ and $f_{n}(T)$ are arbitrary functions of $R$ and $T$ respectively. Each choice of these functional forms leads to distinct dynamical equations and give rise to different cosmological interpretations. In present work, we consider a specific form $f(R,T)=R+f(T)$, where the term $R$ represents the standard Einstein--Hilbert gravitational sector, while the additional correction $f(T)$ represents a modification to GR. This modification introduces additional matter--geometry coupling terms into the gravitational field equations, thereby altering the cosmic dynamics. Such corrections can generate an effective gravitational contribution capable of influencing the late-time expansion history of the universe. Therefore, the Eq.~(4) for the choice $f(R,T)= R+f(T)$ reduces to
\begin{equation}
R_{\mu\nu} - \frac{1}{2}Rg_{\mu\nu} = T_{\mu\nu} - ( T_{\mu\nu}+ \Theta_{\mu\nu})f '(T) + \frac{1}{2}f(T) g_{\mu\nu},
\end{equation}
where the prime denotes the differentiation w.r.t the corresponding argument. The matter Lagrangian $\mathcal L_{m}$ is chosen as $\mathcal  L_{m}=-p$, where $p$ is the thermodynamic pressure of matter content of the universe. Consequently, Eq.~(5) gives $ \Theta_{\mu\nu} = -2 T_{\mu\nu} -p g_{\mu\nu}$. Using this relation, Eq.~(6) reduces to
	\begin{equation}
		R_{\mu\nu} - \frac{1}{2}Rg_{\mu\nu} =  T_{\mu\nu} - ( T_{\mu\nu}+ pg_{\mu\nu})f '(T) + \frac{1}{2}f(T) g_{\mu\nu}.
	\end{equation}
	
In this work, we consider a spatially homogeneous and isotropic flat Friedmann-Lemaitre-Robertson-Walker (FLRW) space-time, which in co-moving coordinates is expressed as
	\begin{equation}
		ds^2 = dt^2 - a^2(t) (dx^2 + dy^2 + dz^2).
	\end{equation}
For the above FLRW line element, Eq.~(8) reduces to the following field equations
	\begin{equation}
		3H^2 = \rho + (\rho + p)f '(T) + \frac{1}{2}f(T),
	\end{equation}
	\begin{equation}
		2\dot{H} + 3H^2 = -p +  \frac{1}{2}f(T).
	\end{equation}	
 Here, $\rho$ and $p$ denote the total energy density and pressure of the cosmic fluid, respectively. In the present work, we adopt the linear function $f(T)=\lambda T$, where $\lambda$ characterizes the strength of the matter--geometry coupling. This simplest non-trivial extension of GR incorporates the effects of matter--geometry coupling while retaining the analytical solvability of the cosmological field equations. Consequently, Eqs.~(9) and (10) reduce to
	\begin{equation}
		3H^{2}=\rho+\lambda(\rho+ p) + \frac{1}{2}\lambda T
	\end{equation}
	\begin{equation}
		2\dot{H} + 3H^{2}= - p +  \frac{1}{2}\lambda T
	\end{equation}
The modified Friedmann Eqs.~(11) and (12) govern the cosmological dynamics of the present $f(R,T)$ gravity model for the adopted linear form. Here, an overdot denotes differentiation with respect to cosmic time. The additional terms involving the coupling parameter $\lambda$ arise from the matter--geometry interaction, thereby modifying the standard cosmological evolution. In the limiting case $\lambda=0$, these equations reduce to the standard Friedmann equations of GR.

The late-time dynamics of the universe are governed predominantly by its dark sector, comprising DM and DE. While modified theories of gravity provide a geometric mechanism for explaining the observed accelerated expansion, combining them with physically motivated dynamical DE models offers a more comprehensive framework for investigating the cosmic expansion history. Among the various dynamical DE scenarios, the running vacuum model has emerged as one of the most promising candidates. Motivated by quantum field theory in curved spacetime, this framework predicts that the vacuum energy density is not a fundamental constant but evolves slowly with the cosmic expansion through quantum corrections. Besides its strong theoretical motivation, the running vacuum scenario has been shown to provide a successful description of the late-time universe while remaining consistent with a wide range of cosmological observations \cite{S2009,So2015,jsola,jsp}.

To investigate the cosmological implications of running vacuum energy in the framework of $f(R,T)$ gravity, we consider a spatially flat universe filled with two dominant components, pressureless DM and running vacuum DE. Since the contribution of baryonic matter to the late-time background evolution is small, it is neglected throughout the present analysis. Consequently, the total energy density of the cosmic fluid is written as $\rho=\rho_m+\rho_\Lambda$, where $\rho_m$ and $\rho_\Lambda$ denote the energy densities of DM and running vacuum energy, respectively. Since dark matter behaves as a pressureless fluid ($p_m=0$), the total pressure of the cosmic fluid is entirely determined by the running vacuum component, i.e., $p=p_\Lambda$. The two components are assumed to interact only through gravity, while the running vacuum energy evolves according to the adopted vacuum law introduced below.

Rather than assuming a constant vacuum energy density, the present framework combines a quantum field theory motivated running vacuum energy density with the matter--geometry coupling inherent in $f(R,T)$ gravity, enabling us to investigate the combined effects of dynamical vacuum evolution and modified gravity on the late-time expansion of the universe. Following the quantum field theory motivated running vacuum framework, we consider the leading late-time contribution to the vacuum energy density in the form \cite{jsp}
\begin{equation}
 \rho_{\Lambda}=\alpha+\beta H^2.
\end{equation}
The above relation implies that the vacuum energy density is no longer strictly constant but evolves with the cosmic expansion. The constant term $\alpha$ represents the residual vacuum energy, whereas the $H^2$ contribution describes the running behaviour of the vacuum at late times. Within the framework of $f(R,T)$ gravity, the matter-geometry coupling additionally modifies the cosmic dynamics, making the present model useful for studying the combined effects of modified gravity and running vacuum energy on the evolution of the universe. Consequently, the Eqs.~(11) and (12) reduces to
\begin{equation}
3H^2 = \rho_{m}+\rho_{\Lambda}+\lambda(\rho_{m}+\rho_{\Lambda}+p_{\Lambda})+\frac{1}{2}\lambda T,
\end{equation}
\begin{equation}
2\dot{H}+3H^2 = -p_{\Lambda}+\frac{1}{2}\lambda T.
\end{equation}

The system contains more unknown quantities than the number of independent field equations and therefore requires an additional relation to obtain a closed system. For the running vacuum component, we assume the equation of state (EoS) $p_{\Lambda}=\omega_{\Lambda}\rho_{\Lambda}$, where $\omega_{\Lambda}$ is the EoS parameter of the running vacuum. Furthermore, the trace of the energy--momentum tensor is given by
$T=\rho_m+\rho_\Lambda-3p_\Lambda$. Using these relations together with the running vacuum energy density, the modified field equations become fully determined. To derive the analytical solution for the Hubble parameter, the explicit expression for $\rho_\Lambda$ is required. Substituting the expression for $\rho_{\Lambda}$ from Eq.~(13) into Eq.~(14) yields
\begin{equation}
	\rho_{m} = \frac{6H^2 - (\alpha + \beta H^2)\left[ 2 + \lambda (3 - \omega_{\Lambda}) \right]}{3\lambda + 2}.
\end{equation}
The above expression shows that the matter energy density is governed by the combined effects of the cosmic expansion, the running vacuum parameters, and the matter--geometry coupling. In particular, the coupling parameter $\lambda$ explicitly characterizes the strength of the matter--geometry interaction and therefore plays a central role in determining the evolution of the matter sector. To determine the Hubble parameter, we combine Eqs.~(14) and (15), which yields
\begin{equation}
	2\dot{H} + (\lambda + 1)\left[ \rho_{m} + (1 + \omega_{\Lambda}) \rho_{\Lambda} \right] = 0.
\end{equation}

 Substituting the expressions for $\rho_\Lambda$ and $\rho_m$ from Eqs.~(13) and (16) into the above equation yields
\begin{equation}
\dot{H}+\frac{(\lambda+1)}{(3\lambda+2)}[3H^2+(\alpha+\beta H^2)(2\lambda+1)\omega_{\Lambda}=0.
\end{equation}
For observational analysis, it is convenient to express the Hubble parameter in terms of the redshift parameter $z$. Using the relation
$\frac{d}{dt}=-(1+z)H\frac{d}{dz}$,
the evolution equation can be rewritten in terms of the redshift parameter, yielding the Hubble parameter in the form $H=H(z)$ as
\begin{equation}
H(z)=\sqrt{C_{0}(1+z)^{\frac{2(\lambda+1)(3+\beta(2\lambda+1)\omega_{\Lambda})}{(3\lambda+2)}}-\frac{\alpha(2\lambda+1)\omega_{\Lambda}}{3+\beta(2\lambda+1)\omega_{\Lambda}}},
\end{equation}
where $C_{0}$ is a positive integration constant. Applying the present-day boundary condition $H(0)=H_{0}$, the above equation reduces to
\begin{equation}
H(z)=
\sqrt{
	\left(
	H_{0}^{2}+\frac{\alpha(2\lambda+1)\omega_{\Lambda}}{3+\beta(2\lambda+1)\omega_{\Lambda}}
	\right)
	(1+z)^{\frac{2(\lambda+1)\left(3+\beta(2\lambda+1)\omega_{\Lambda}\right)}{(3\lambda+2)}}
	-
	\frac{\alpha(2\lambda+1)\omega_{\Lambda}}{3+\beta(2\lambda+1)\omega_{\Lambda}}.
}
\end{equation}
This exact analytical expression forms the basis for the subsequent observational analysis and the reconstruction of various cosmological parameters.

\section{Observational Data Sets}
To assess the observational viability of the proposed running vacuum cosmology in the framework of $f(R,T)$ gravity, we employ recent late-time cosmological observations comprising the Pantheon+SH0ES Type Ia supernova compilation and the latest BAO measurements from DESI. These complementary datasets provide stringent constraints on the expansion history of the universe over a broad redshift range \cite{Brout2022,DESI2024}. The model is characterized by the parameter set
\begin{equation}
	\Theta=\{\alpha,\beta,\lambda,\omega_{\Lambda},H_0\},
\end{equation}
where $\alpha$ and $\beta$ describe the running vacuum energy density, $\lambda$ denotes the matter--geometry coupling parameter in $f(R,T)$ gravity, $\omega_{\Lambda}$ is the dark energy equation-of-state parameter, and $H_0$ represents the present-day Hubble constant.

\subsection{Pantheon+SH0ES Type Ia Supernovae}

The primary observational constraint is provided by the Pantheon+SH0ES compilation \cite{Brout2022}, which consists of 1701 spectroscopically confirmed Type Ia supernovae spanning the redshift interval $0.001<z<2.3$. The compilation combines the Pantheon sample with the SH0ES Cepheid calibration. It currently represents one of the most comprehensive and precise compilations for constraining the late-time expansion history of the universe. The observed distance modulus is defined as
\begin{equation}
	\mu_{\rm obs}=m_B-M,
\end{equation}
where $m_B$ is the apparent magnitude and $M$ denotes the absolute magnitude of Type Ia supernovae. The corresponding theoretical distance modulus is given by
\begin{equation}
	\mu_{\rm th}(z)=
	5\log_{10}\left[D_L(z)\right]+25,
\end{equation}
where the luminosity distance is
\begin{equation}
	D_L(z)=(1+z)c\int_0^z \frac{dz'}{H(z')}.
\end{equation}
The complete covariance matrix, including both statistical and systematic uncertainties, is incorporated into the analysis. Since the Pantheon+SH0ES compilation is calibrated using the SH0ES Cepheid distance ladder, the Hubble constant calibration is already incorporated and the absolute magnitude is therefore not introduced as an additional nuisance parameter. The corresponding chi-square statistic is
\begin{equation}
	\chi^2_{\rm SN}
	=
	\Delta\mu^{\rm T}
	C^{-1}
	\Delta\mu,
\end{equation}
where
\[
\Delta\mu=
\mu_{\rm obs}-\mu_{\rm th},
\]
and $C$ denotes the full covariance matrix of the Pantheon+SH0ES sample.

\subsection{DESI BAO Measurements}

To complement the supernova observations, we incorporate the latest BAO measurements released by the DESI collaboration \cite{DESI2024}. The DESI dataset provides precise BAO measurements over multiple redshift bins, thereby offering complementary constraints on the cosmic expansion history. The BAO observables are expressed as
\begin{equation}
	\frac{D_M(z)}{r_d},
	\qquad
	\frac{D_H(z)}{r_d},
\end{equation}
where
\begin{equation}
	D_M(z)=
	c\int_0^z\frac{dz'}{H(z')}
\end{equation}
is the comoving angular diameter distance and
\begin{equation}
	D_H(z)=\frac{c}{H(z)}
\end{equation}
is the Hubble distance. Here, $r_d$ denotes the comoving sound horizon at the drag epoch. Following the standard late-time cosmological analyses, the sound horizon at the drag epoch is fixed to $r_d=147~{\rm Mpc}$. The DESI likelihood is constructed using the full covariance matrix provided by the collaboration, and the corresponding chi-square function is written as
\begin{equation}
	\chi^2_{\rm BAO}
	=
	\Delta F^{\rm T}
	C_{\rm BAO}^{-1}
	\Delta F,
\end{equation}
where $\Delta F$ denotes the difference between the theoretical predictions and the observed BAO measurements.

\subsection{Joint Analysis}
To obtain robust constraints on the cosmological parameters, we perform a joint likelihood analysis by combining the Pantheon+SH0ES and DESI datasets. In addition to the individual analyses using each dataset separately, we also consider their combined dataset to exploit their complementary constraining power. Assuming that the two datasets are statistically independent, the total likelihood is constructed from the sum of the individual chi-square functions,
\begin{equation}
	\chi^2_{\rm total}
	=
	\chi^2_{\rm SN}
	+
	\chi^2_{\rm BAO}.
\end{equation}

The posterior probability distributions of the model parameters are explored using the affine-invariant Markov Chain Monte Carlo (MCMC) ensemble sampler implemented in the \texttt{emcee} package. The statistical analysis is performed in the five-dimensional parameter space and the resulting posterior distributions are used to determine the best-fit values together with their corresponding confidence intervals. The observational constraints obtained from the individual and joint analyses are presented in the following section.
\section{Observational Constraints and Best-Fit Analysis}

The free parameters of the proposed cosmological model are constrained using the likelihood functions constructed in the previous section. The parameter estimation is performed through the Markov Chain Monte Carlo (MCMC) technique employing the \texttt{emcee} package with the affine-invariant ensemble sampler. To ensure reliable parameter estimation, the Markov chains are evolved for a sufficiently large number of sampling steps, and the initial burn-in phase is discarded before constructing the final posterior distributions.

\begin{figure}
	\centering
	\includegraphics[width=0.75\textwidth]{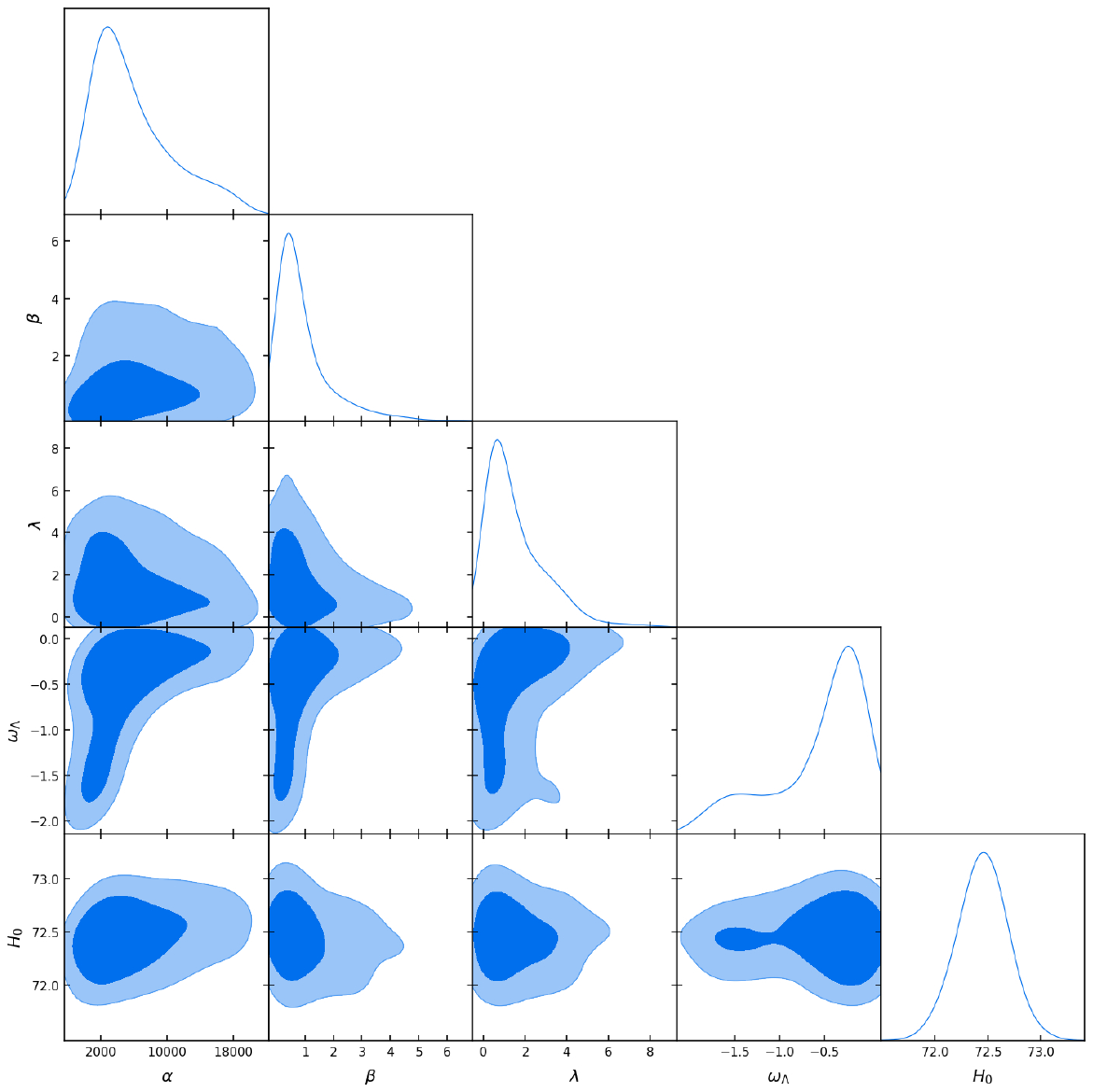}
	\caption{MCMC confidence contours for the Pantheon+SH0ES dataset.}
	\label{fig:pantheon_contour}
\end{figure}

\begin{figure}
	\centering
	\includegraphics[width=0.75\textwidth]{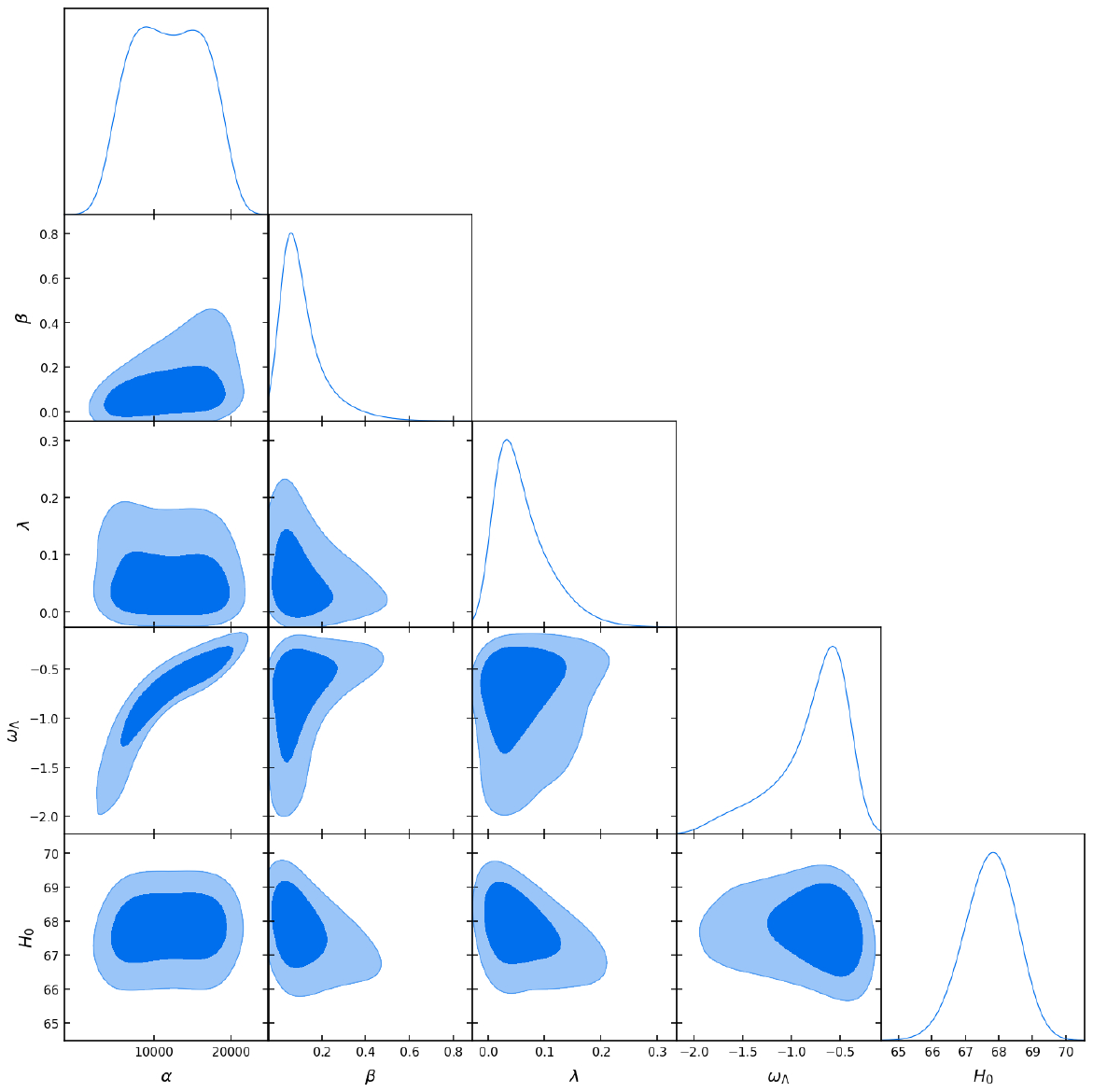}
	\caption{MCMC confidence contours obtained using the DESI BAO dataset.}
	\label{fig:desi_contour}
\end{figure}

\begin{figure}
	\centering
	\includegraphics[width=0.75\textwidth]{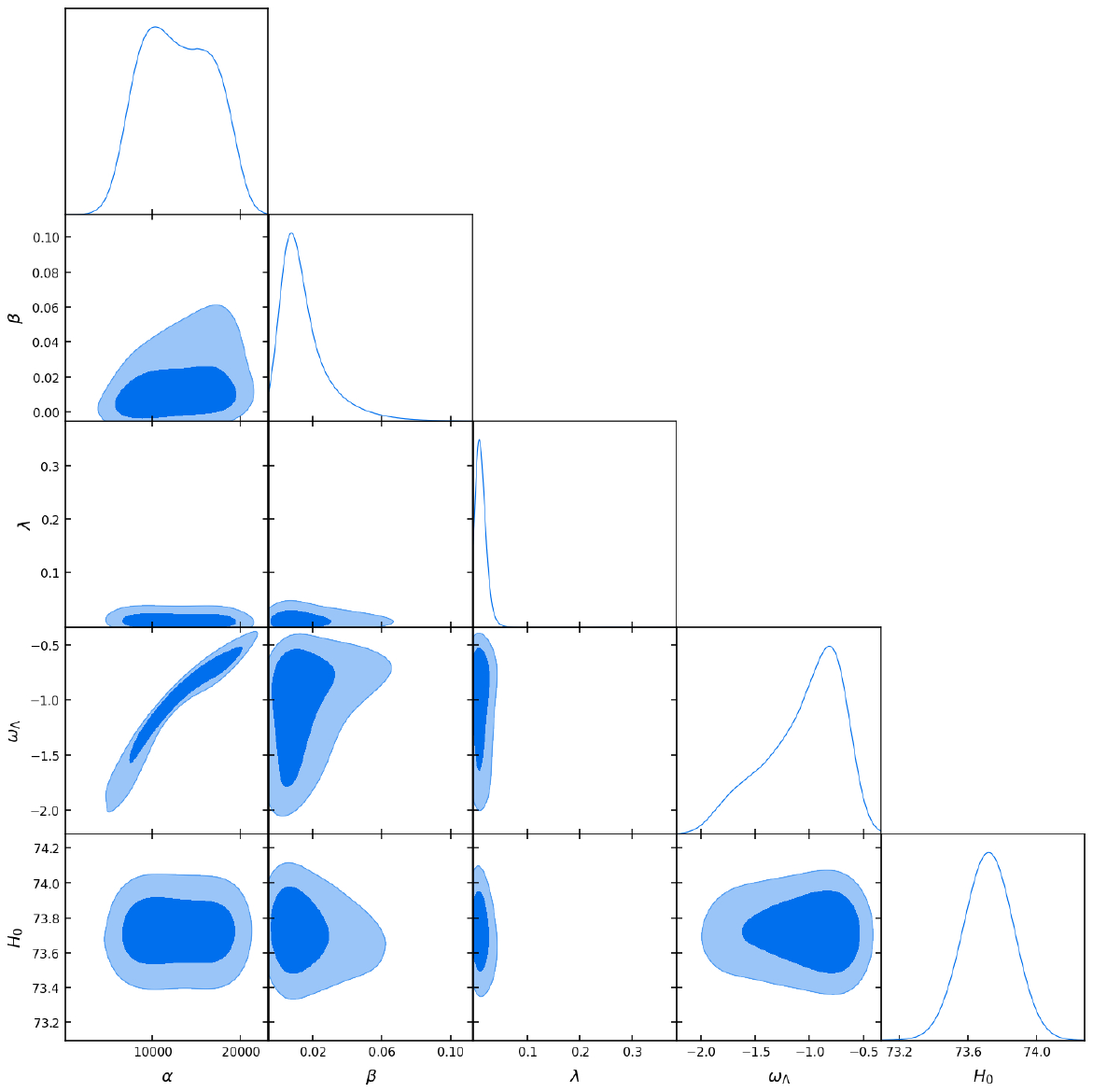}
	\caption{Joint confidence contours for the combined Pantheon+SH0ES+DESI dataset.}
	\label{fig:combined_contour}
\end{figure}

The resulting marginalized one and two-dimensional posterior distributions for the Pantheon+SH0ES, DESI, and the combined Pantheon+SH0ES+DESI datasets are presented in Figs.~\ref{fig:pantheon_contour}, \ref{fig:desi_contour}, and \ref{fig:combined_contour}, respectively. These contour plots illustrate the confidence regions and correlations among the model parameters. The corresponding best-fit values together with their $1\sigma$ confidence intervals are summarized in Table~\ref{tab:bestfit}. The obtained values of $\chi^2/\mathrm{dof}$ remain close to unity, indicating good agreement between the theoretical predictions and the observational data. 

Among the three cases, the combined Pantheon+SH0ES+DESI dataset yields the tightest constraints on the model parameters, reflecting the complementary constraining power of Type Ia supernovae and BAO observations and leading to more precise estimates of the cosmological parameters. A notable feature of the combined analysis is the significant reduction in the uncertainties of the running vacuum parameter $\beta$ and the matter--geometry coupling parameter $\lambda$. The preferred value of $\beta$ remains small but positive, indicating that the present observations allow only mild vacuum dynamics while remaining compatible with the standard late-time expansion history. Similarly, the matter--geometry coupling parameter $\lambda$ is constrained to values close to zero, suggesting that any deviation from GR through the matter--geometry coupling is tightly limited by current observations. Furthermore, the EoS parameter approaches $\omega_{\Lambda}\approx-1$, indicating that the effective DE component exhibits a nearly cosmological-constant-like behaviour in the combined analysis.

The reconstructed values of the Hubble constant also exhibit an interesting trend. While the Pantheon+SH0ES dataset favours a comparatively larger value of $H_0$, the DESI-only analysis yields a lower estimate that is closer to the value inferred from early-universe observations. The combined analysis favours a comparatively higher value of $H_0$ than the individual Pantheon+SH0ES and DESI analyses, remaining consistent with late-time distance-ladder measurements. The comparatively lower value of $H_0$ obtained from the DESI-only analysis reflects the preference of BAO measurements for an expansion history that is broadly consistent with early-universe constraints. The inclusion of the Pantheon+SH0ES dataset shifts the preferred value towards higher $H_0$, highlighting the complementary constraining power of the two observational probes and yielding a joint estimate consistent with late-time distance-ladder measurements.
 
The obtained observational constraints are broadly consistent with previous studies of running vacuum cosmologies, where the vacuum dynamics is generally found to be weak and characterized by small deviations from a constant vacuum energy density \cite{jsola}. In particular, the small preferred values of the running vacuum parameter and the matter--geometry coupling parameter are also in qualitative agreement with the recent observational analysis of running vacuum models in $f(R,T)$ gravity reported by Errahmani \textit{et al.} \cite{Errahmani}, where the parameters distinguishing the running vacuum scenario from the standard cosmological model were likewise found to be tightly constrained. 

Overall, the observational constraints indicate that the proposed running vacuum model in $f(R,T)$ gravity remains consistent with current cosmological observations and provides a viable description of the late-time expansion history of the universe.

\begin{table*}[t]
	\centering
	\caption{Best-fit values of model parameters obtained from different observational datasets.}
	\label{tab:bestfit}
	\begin{tabular}{|c|c|c|c|}
		\hline
		Parameter & Pantheon+SH0ES & DESI & Pantheon+SH0ES+DESI \\
		\hline
		$\alpha$
		& $5670.22599 \pm 4916.74901$
		& $11956.43568 \pm 4614.80486$
		& $12827.88334 \pm 3972.01785$ \\
		
		$\beta$
		& $0.85666 \pm 0.99987$
		& $0.10856 \pm 0.11170$
		& $0.01428 \pm 0.01503$ \\
		
		$\lambda$
		& $1.48331 \pm 1.48888$
		& $0.05931 \pm 0.05002$
		& $0.01088 \pm 0.02296$ \\
		
		$\omega_{\Lambda}$
		& $-0.54182 \pm 0.53325$
		& $-0.77297 \pm 0.37671$
		& $-1.04009 \pm 0.35547$ \\
		
		$H_0$
		& $72.45027 \pm 0.24575$
		& $67.74745 \pm 0.78955$
		& $73.71868 \pm 0.14707$ \\
		\hline
		
		$\chi^2$
		& $1748.9655$
		& $9.7099$
		& $1918.7571$ \\
		
		$\chi^2/\mathrm{dof}$
		& $1.0312$
		& $1.2137$
		& $1.1227$ \\
		\hline
	\end{tabular}
\end{table*}

\section{Cosmological Evolution}
In this section, we investigate the cosmological implications of the present running vacuum model using the best-fit values obtained from the observational analysis. In particular, we study the evolution of the Hubble parameter, the deceleration parameter, the effective EoS parameter, the age of the universe, and the consistency of the model with observational Hubble data through residual analysis. The reconstructed cosmological behavior is analyzed using the best-fit parameter values obtained from the Pantheon+SH0ES, DESI, and combined Pantheon+SH0ES+DESI datasets.
\subsection{Evolution of the Hubble Parameter}
The Hubble parameter $H(z)$ describes the expansion rate of the universe at different cosmic epochs. Using the constrained model parameters, we reconstruct the evolution of the Hubble parameter and compare it with observational Hubble data as well as the standard $\Lambda$CDM cosmology.

\begin{figure}
	\centering
	\includegraphics[width=0.8\linewidth]{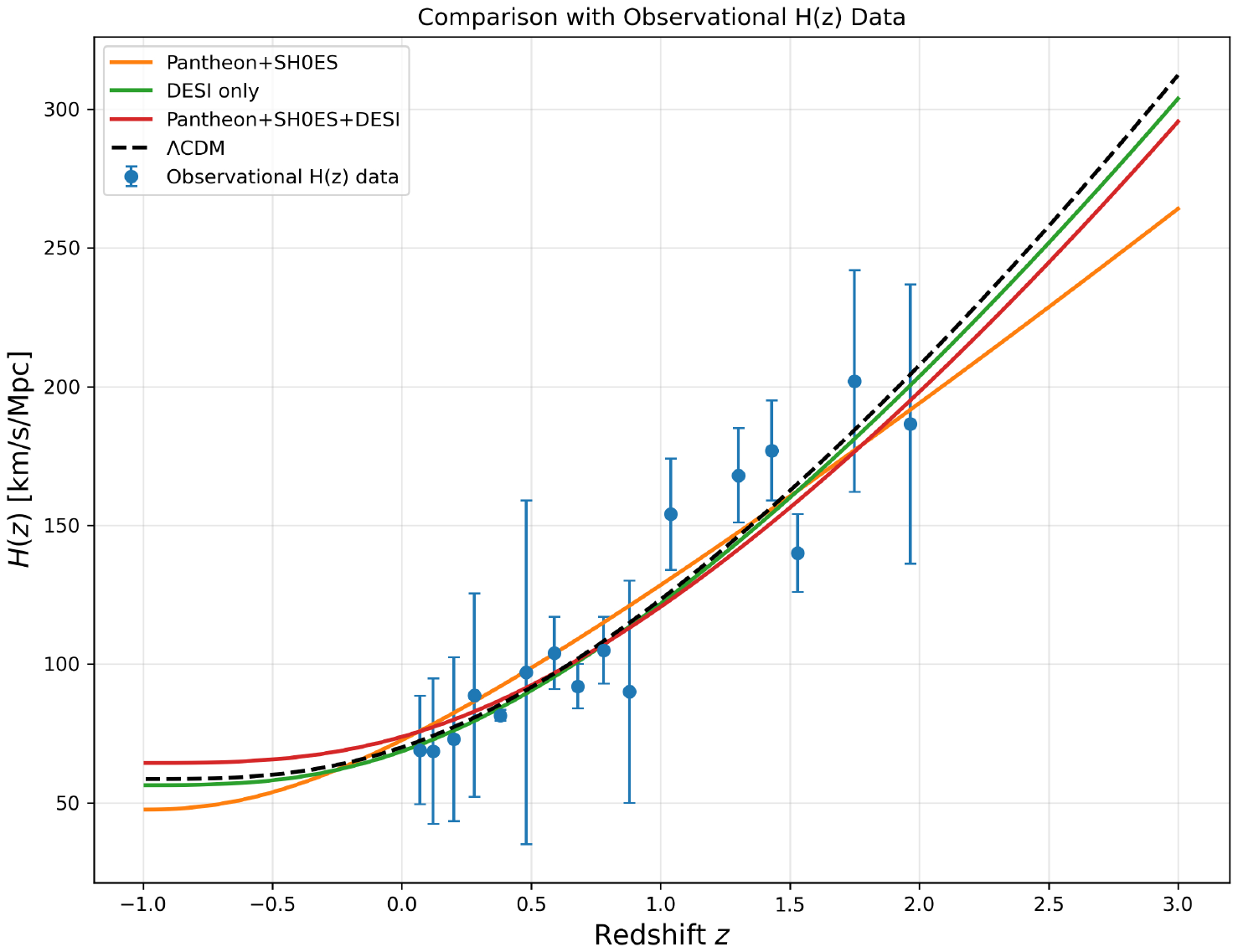}
	\caption{Comparison of the reconstructed Hubble parameter $H(z)$ with observational Hubble data for different datasets. The dashed black curve corresponds to the standard $\Lambda$CDM model.}
	\label{fig:Hz_plot}
\end{figure}

Figure~\ref{fig:Hz_plot} presents the reconstructed evolution of the Hubble parameter obtained from the three observational analyses together with the observational $H(z)$ data and the standard $\Lambda$CDM prediction. All reconstructed curves exhibit the expected increase with redshift and remain consistent with the observational measurements within their uncertainties. At low redshifts, the three reconstructions show only minor differences, whereas noticeable deviations appear towards higher redshifts. In particular, the Pantheon+SH0ES reconstruction predicts comparatively smaller values of $H(z)$, while the DESI-only and combined Pantheon+SH0ES+DESI reconstructions closely follow the $\Lambda$CDM expansion history over the entire redshift range considered. This behaviour indicates that the observationally preferred running vacuum model reproduces the standard late-time expansion history while allowing only mild departures from the $\Lambda$CDM scenario.

\subsection{Deceleration Parameter}
The deceleration parameter $q(z)$ is defined as in terms of redshift $q(z) = -1 + (1+z)\frac{1}{H(z)}\frac{dH(z)}{dz}$ and derived  as
\begin{equation}
	q(z)
	=
	-1+\frac{(1+z)}{H(z)}
	\frac{dH(z)}{dz}.
\end{equation}
Using the reconstructed Hubble parameter,
\begin{equation}
	H^2(z)=\left(H_0^2+A\right)(1+z)^n-A,
\end{equation}
where
\begin{equation}
	A=
	\frac{\alpha(1+2\lambda)\omega_\Lambda}
	{3+\beta\omega_\Lambda+2\beta\lambda\omega_\Lambda},
\end{equation}
and
\begin{equation}
	n=
	\frac{2(1+\lambda)\left(3+\beta\omega_\Lambda+2\beta\lambda\omega_\Lambda\right)}
	{2+3\lambda},
\end{equation}
we analytically obtain
\begin{equation}
	\frac{dH}{dz}
	=
	\frac{
		n(H_0^2+A)(1+z)^{n-1}
	}
	{2H(z)}.
\end{equation}
Consequently, the deceleration parameter becomes
\begin{equation}
	q(z)
	=
	-1+
	\frac{
		n(H_0^2+A)(1+z)^n
	}
	{2H^2(z)}.
\end{equation}
At the present epoch $(z=0)$, the present deceleration parameter is given by
\begin{equation}
	q_0
	=
	-1+
	\frac{
		n(H_0^2+A)
	}
	{2H_0^2}.
\end{equation}

\begin{figure}
	\centering
	\includegraphics[width=0.8\linewidth]{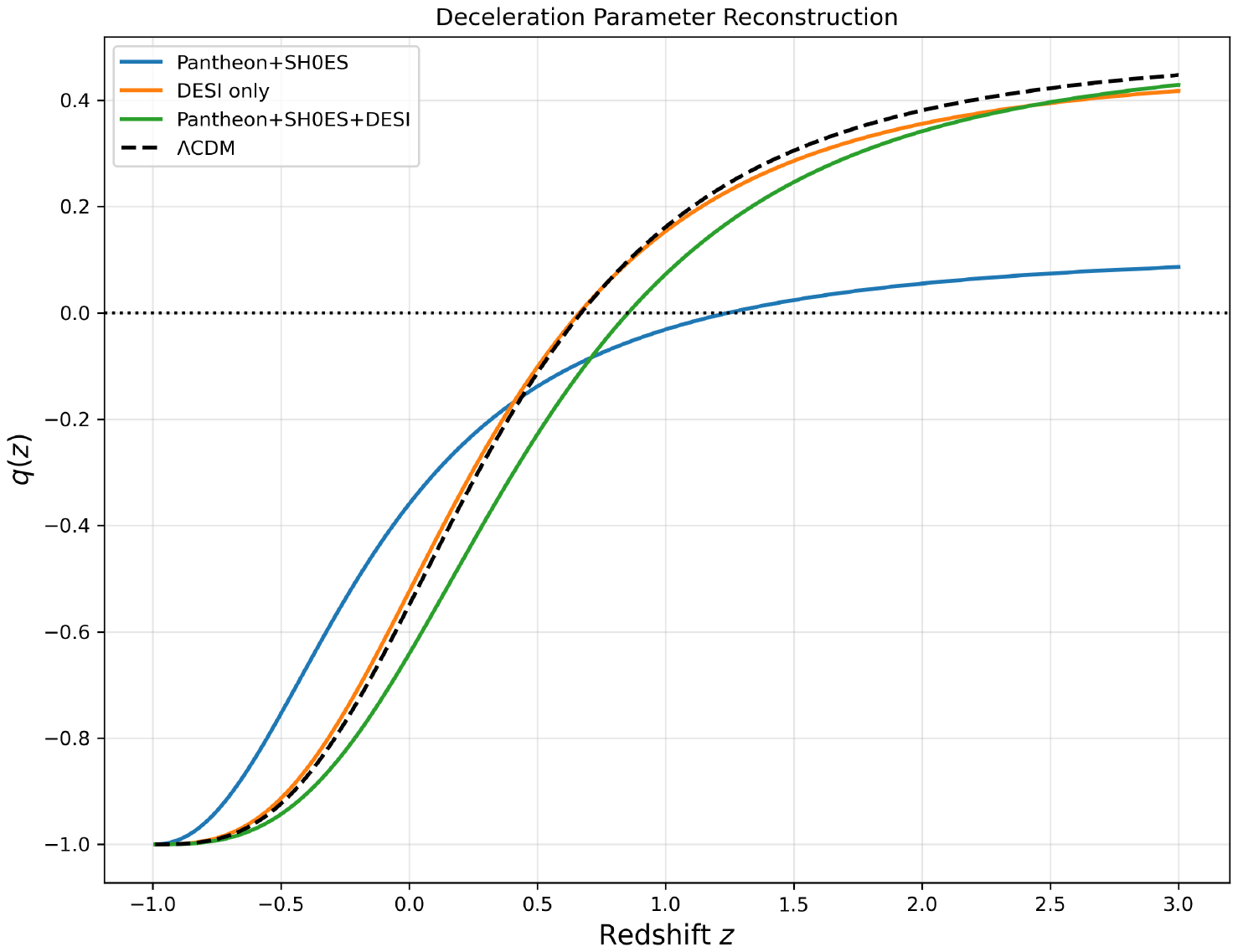}
	\caption{Evolution of the deceleration parameter $q(z)$ for different observational datasets. The transition from decelerated expansion to accelerated expansion is clearly visible.}
	\label{fig:qz_plot}
\end{figure}

Figure~\ref{fig:qz_plot} illustrates the reconstructed evolution of the DP obtained from the Pantheon+SH0ES, DESI, and combined Pantheon+SH0ES+DESI analyses. At high redshifts, all reconstructed curves attain positive values of $q(z)$, indicating that the universe experienced a decelerating expansion during the matter-dominated epoch. As the universe evolves, the DP decreases and eventually becomes negative, explaining the present accelerated expansion. The reconstructed present values are found to be approximately $q_0\approx-0.36$ (Pantheon+SH0ES), $q_0\approx-0.53$ (DESI), and $q_0\approx-0.64$ (Pantheon+SH0ES+DESI). These results indicate that the DESI and joint analyses favour a comparatively stronger present-day cosmic acceleration than the Pantheon+SH0ES dataset alone. In particular, the combined reconstruction follows the $\Lambda$CDM prediction over most of the redshift range while allowing a slightly stronger late-time acceleration, consistent with the observational preference for an effective EoS parameter approaching $\omega_\Lambda\simeq-1$ obtained from the joint parameter estimation.

The transition redshift, determined from the condition $q(z_t)=0$, lies in the interval $0.6\lesssim z_t\lesssim1.3$, depending on the observational dataset employed.  This range is broadly consistent with recent observational estimates of the cosmic acceleration transition. Although the Pantheon+SH0ES reconstruction predicts the earliest transition to accelerated expansion, it yields the least negative present-day DP. This behaviour reflects a comparatively milder acceleration history after the transition. In contrast, the DESI and joint analyses favour a slightly later transition but a more rapid evolution towards the present accelerated epoch, resulting in larger negative values of the present DP. The combined Pantheon+SH0ES+DESI reconstruction yields a transition history that most closely follows the standard cosmological scenario while retaining the dynamical character of the running vacuum model. The reconstructed evolution of the DP therefore indicates that the proposed running vacuum model in $f(R,T)$ gravity successfully captures the observed transition from an early decelerating phase to the present accelerated expansion while preserving the standard background expansion history.

\subsection{Effective Equation of State Parameter}
The effective EoS parameter is defined through the relation
\begin{equation}
	\omega_{\rm eff}(z)
	=
	\frac{2q(z)-1}{3}.
\end{equation}

\begin{figure}
	\centering
	\includegraphics[width=0.8\linewidth]{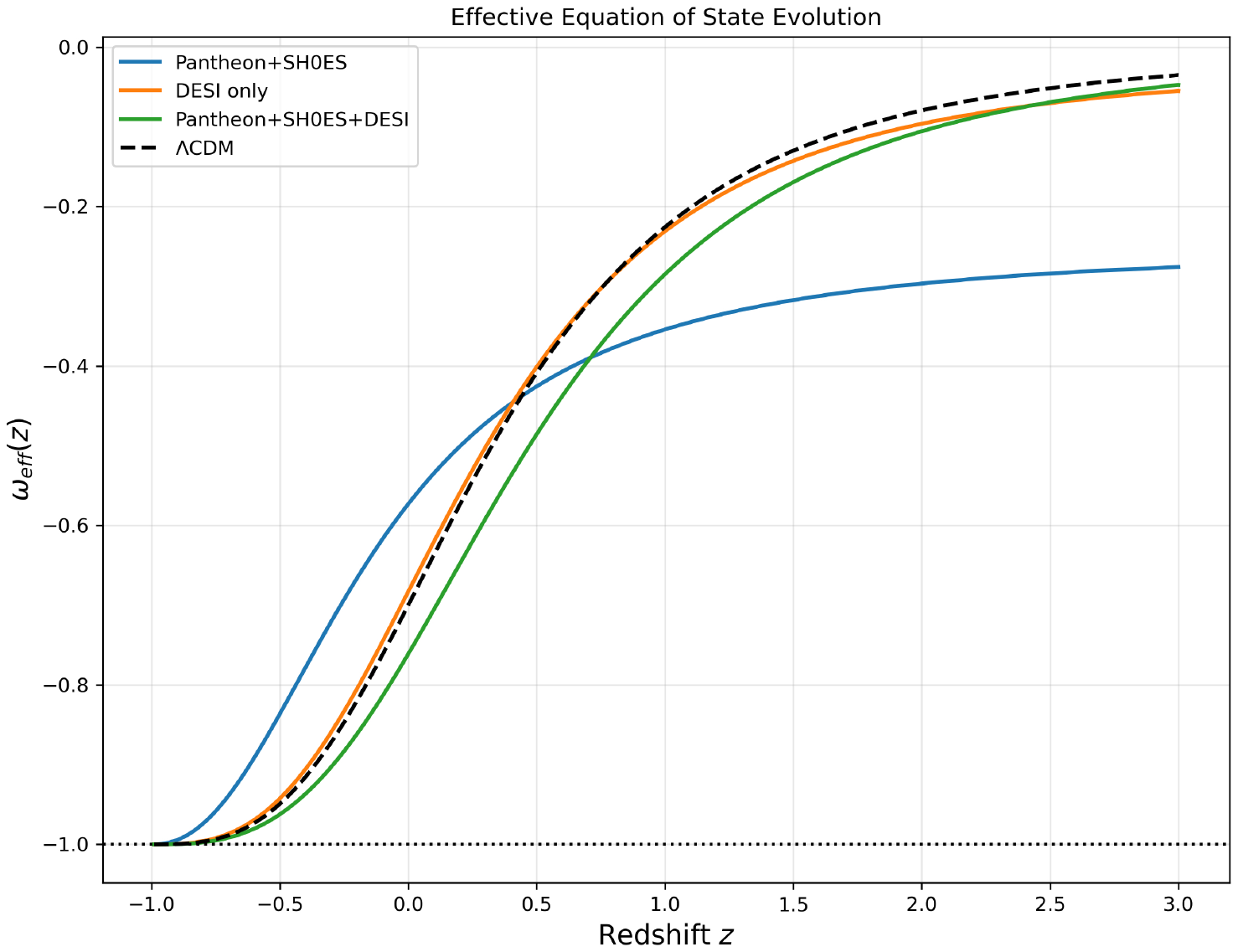}
	\caption{Evolution of the effective equation-of-state parameter reconstructed from the Pantheon+SH0ES, DESI, and joint Pantheon+SH0ES+DESI analyses. The horizontal dotted line represents the cosmological constant boundary $(\omega=-1)$.}
	\label{fig:eos_plot}
\end{figure}

Figure~\ref{fig:eos_plot} presents the reconstructed evolution of the effective EoS parameter obtained from the Pantheon+SH0ES, DESI, and combined Pantheon+SH0ES+DESI analyses. At high redshifts, all reconstructed curves approach $\omega_{\rm eff}\simeq0$, indicating that the cosmic expansion is governed predominantly by pressureless matter. As the universe evolves, $\omega_{\rm eff}$ gradually decreases towards negative values, reflecting the increasing contribution of the running vacuum component to the cosmic energy budget and the onset of late-time accelerated expansion.

The reconstructed present values are approximately $\omega_{\rm eff,0}\approx-0.57$ (Pantheon+SH0ES), $\omega_{\rm eff,0}\approx-0.70$ (DESI), and $\omega_{\rm eff,0}\approx-0.73$ (Pantheon+SH0ES+DESI). Since all reconstructed values satisfy $\omega_{\rm eff,0}<-\frac{1}{3}$, the model consistently predicts the present accelerated expansion of the universe. Among the three analyses, the Pantheon+SH0ES dataset yields the least negative present value of $\omega_{\rm eff}$, indicating a comparatively weaker effective negative pressure. In contrast, the DESI and joint analyses favour more negative values of $\omega_{\rm eff}$, corresponding to a stronger effective repulsive component responsible for the current cosmic acceleration.

The joint Pantheon+SH0ES+DESI reconstruction closely follows the $\Lambda$CDM prediction over the entire redshift range while retaining a mildly dynamical behaviour associated with the running vacuum framework. This behaviour is consistent with the observational constraints obtained in the previous section, where the joint analysis favours an EoS parameter close to $\omega_\Lambda=-1$ together with a small positive running vacuum parameter. Overall, the reconstructed effective EoS evolution demonstrates that the proposed running vacuum model in $f(R,T)$ gravity provides a realistic description of the transition from a matter-dominated universe to the present epoch of accelerated expansion while remaining compatible with current cosmological observations.

\subsection{Age of the Universe}
The age of the universe at a given redshift is defined by
\begin{equation}
	t(z)
	=
	\frac{1}{H_0}
	\int_z^\infty
	\frac{dz'}
	{(1+z')E(z')},
\end{equation}
where $ E(z)=\frac{H(z)}{H_0}$.

\begin{figure}
	\centering
	\includegraphics[width=0.8\linewidth]{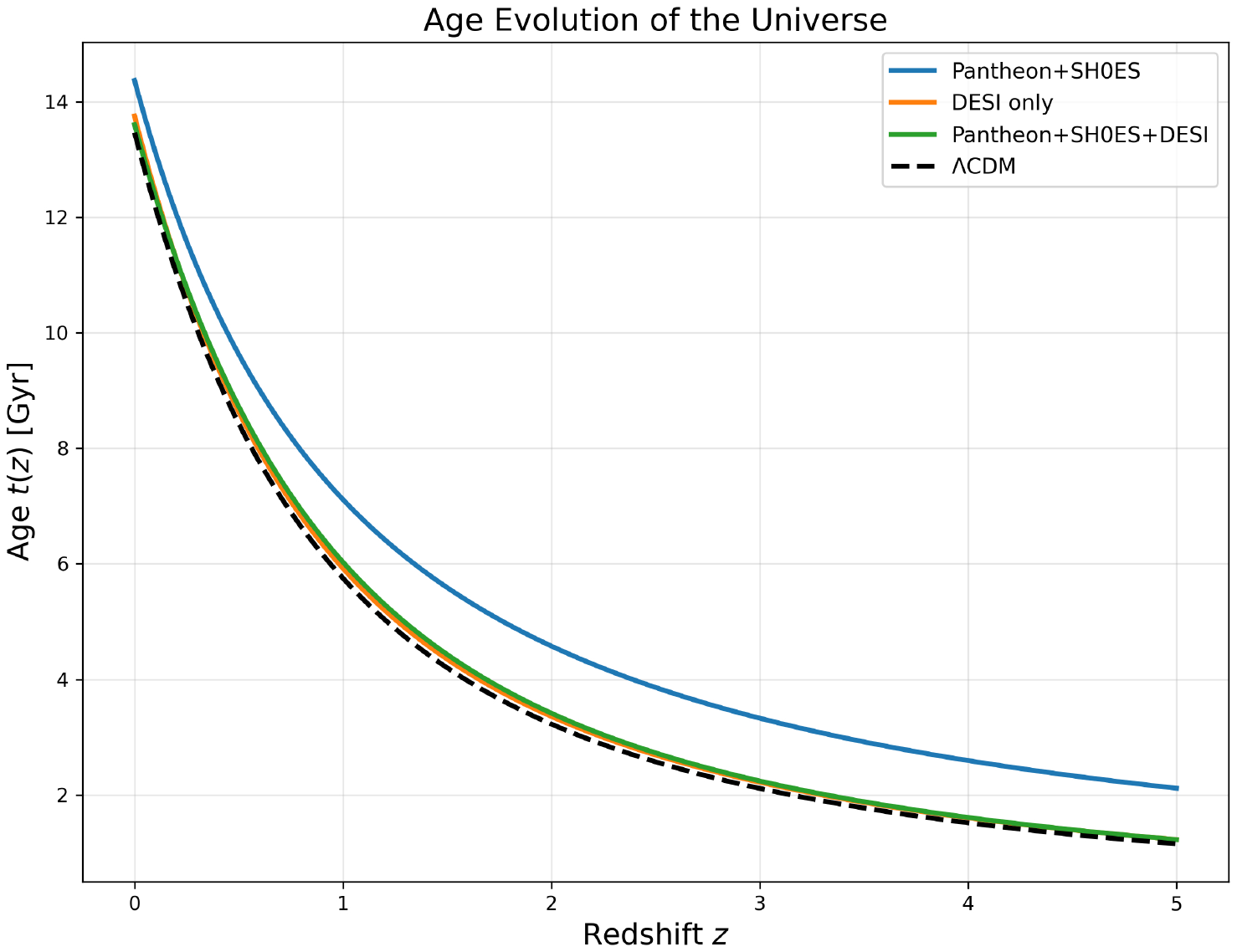}
	\caption{Evolution of the age of the Universe with redshift for different observational datasets.}
	\label{fig:age_plot}
\end{figure}

Using the best-fit parameter values, we reconstruct the age evolution of the universe for the Pantheon+SH0ES, DESI, and combined Pantheon+SH0ES+DESI datasets. The corresponding present ages are found to be $t_0=14.37~{\rm Gyr}$ (Pantheon+SH0ES), $13.75~{\rm Gyr}$ (DESI), and $13.60~{\rm Gyr}$ (Pantheon+SH0ES+DESI), while the standard $\Lambda$CDM model predicts $t_0\approx13.47~{\rm Gyr}$. Figure~\ref{fig:age_plot} illustrates the evolution of the cosmic age as a function of redshift. As expected, the age decreases monotonically with increasing redshift, reflecting the progressively younger universe at earlier cosmic epochs. The DESI and joint Pantheon+SH0ES+DESI reconstructions remain very close to the $\Lambda$CDM prediction throughout the entire redshift range, indicating that the proposed running vacuum model preserves the standard cosmological timescale. Although the Pantheon+SH0ES reconstruction predicts a comparatively older universe, its estimated age remains compatible with current observational estimate of the age of the universe ($13.80~{\rm Gyr}$) \cite{Planck}.

Overall, the reconstructed age evolution demonstrates that the proposed running vacuum model in $f(R,T)$ gravity provides a realistic description of the cosmic expansion history while remaining consistent with the observationally inferred age of the universe.

\subsection{Validation with Cosmic Chronometer Data}

To further assess the predictive capability of the proposed model, we compare the reconstructed Hubble parameter with an independent compilation of Cosmic Chronometer (CC) measurements through a normalized residual analysis. Since the CC data are not included in the parameter estimation, this comparison provides an independent validation of the observationally constrained model. The normalized residual is defined as
\begin{equation}
	R(z)=\frac{H_{\rm CC}(z)-H_{\rm model}(z)}{\sigma_H},
\end{equation}
where $H_{\rm CC}$ denotes the observed Hubble parameter obtained from Cosmic Chronometer measurements, $H_{\rm model}$ is the corresponding theoretical prediction, and $\sigma_H$ represents the observational uncertainty.

\begin{figure}
	\centering
	\includegraphics[width=0.8\linewidth]{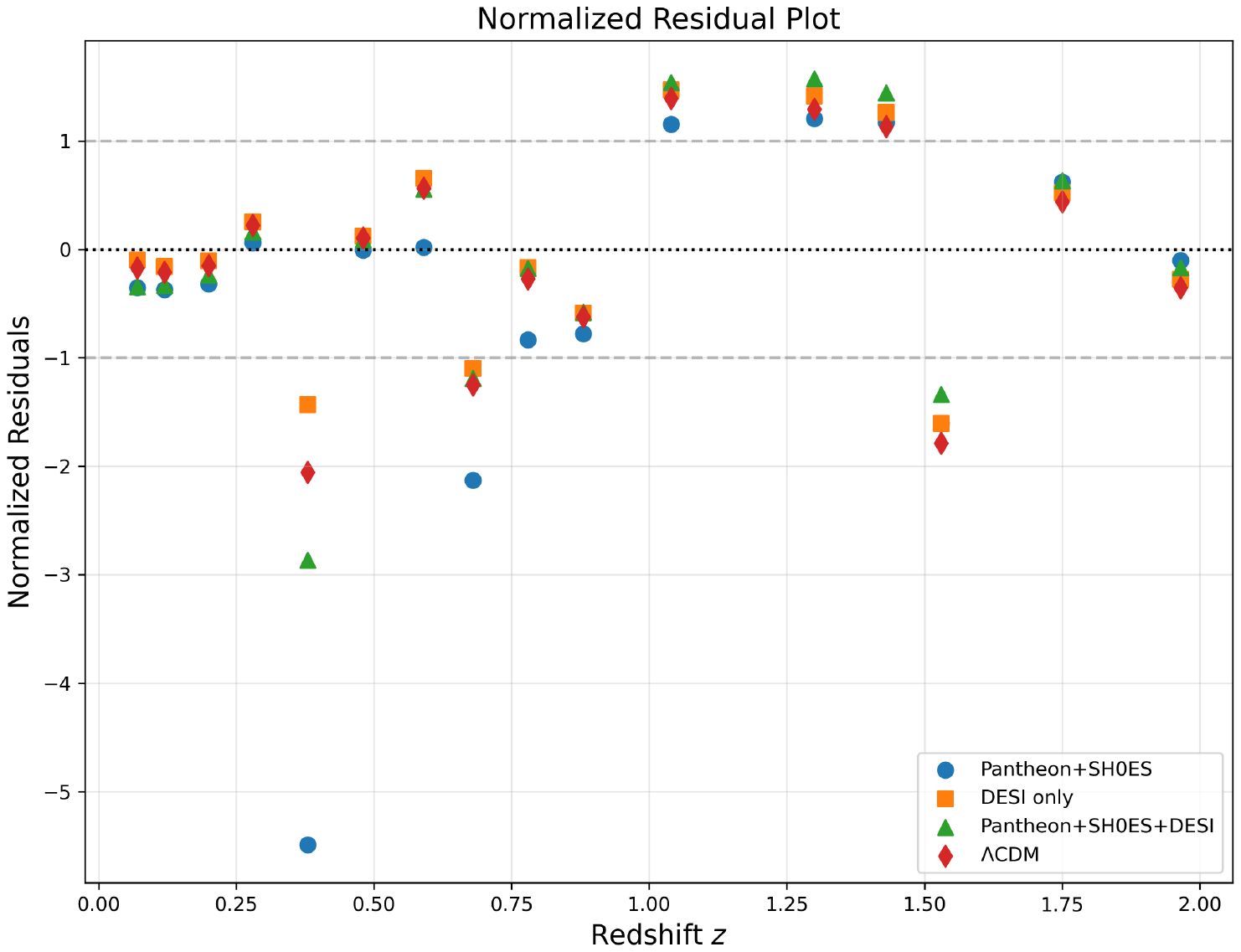}
	\caption{Normalized residuals between the reconstructed Hubble parameter and observational $H(z)$ measurements for different datasets.}
	\label{fig:residual_plot}
\end{figure}

Figure~\ref{fig:residual_plot} presents the normalized residuals corresponding to the Pantheon+SH0ES, DESI, combined Pantheon+SH0ES+DESI reconstructions, and the reference $\Lambda$CDM model. Overall, the residuals are distributed around the zero line, with the majority of the measurements satisfying $|R|\lesssim2$, indicating satisfactory agreement between the reconstructed expansion histories and the independent CC observations. A notable feature of the residual distribution is the presence of a localized negative deviation within the intermediate-redshift interval $0.30\lesssim z\lesssim0.45$. Importantly, the residuals return close to zero at the neighbouring CC measurements, indicating that the observed discrepancy is confined to a narrow redshift interval rather than representing a systematic offset across the entire expansion history. This localized behaviour is evident in the Pantheon+SH0ES, DESI, and combined Pantheon+SH0ES+DESI reconstructions, as well as in the reference $\Lambda$CDM model, although the magnitude of the residuals differs among them. Interestingly, this residual feature lies within the broader intermediate-redshift regime where recent model-independent reconstruction studies have also reported a localized departure from the Planck $\Lambda$CDM expansion history. Although the present analysis is based on an independent CC residual test rather than a reconstruction-based tension statistic, the qualitative overlap in the location of the maximum deviation suggests that the intermediate-redshift region deserves further investigation with future high-precision cosmological observations~\cite{Cho}.

The residual analysis demonstrates that the proposed running vacuum model reproduces the independent CC measurements with an accuracy comparable to that of the standard $\Lambda$CDM model. Since a similar localized residual feature is also present in the reference $\Lambda$CDM model, the observed discrepancy cannot be regarded as a deficiency specific to the proposed cosmological model. Instead, it may reflect characteristics of the intermediate-redshift observational window or subtle features of the late-time expansion history. Therefore, the residual analysis provides no evidence for a systematic inadequacy of the proposed model and further supports its consistency with independent CC observations while emphasizing the importance of the intermediate-redshift regime for future observational studies.
\section{Diagnostic Analysis}
Although the Hubble parameter and the DP provide valuable information about the expansion history of the universe, they are often insufficient to distinguish between cosmological models exhibiting similar background evolution. To overcome this limitation, higher-order geometrical diagnostics have been introduced as effective tools for characterizing the dynamical nature of DE. Among these, the statefinder $(r,s)$ pair and the Om diagnostic are particularly useful, since they can efficiently discriminate the standard $\Lambda$CDM model from alternative DE scenarios. In this section, we investigate these diagnostics using the best-fit parameters obtained from the Pantheon+SH0ES, DESI, and the combined Pantheon+SH0ES+DESI datasets.
\subsection{Statefinder Diagnostic}
\begin{figure}
	\centering
	\includegraphics[width=0.75\textwidth]{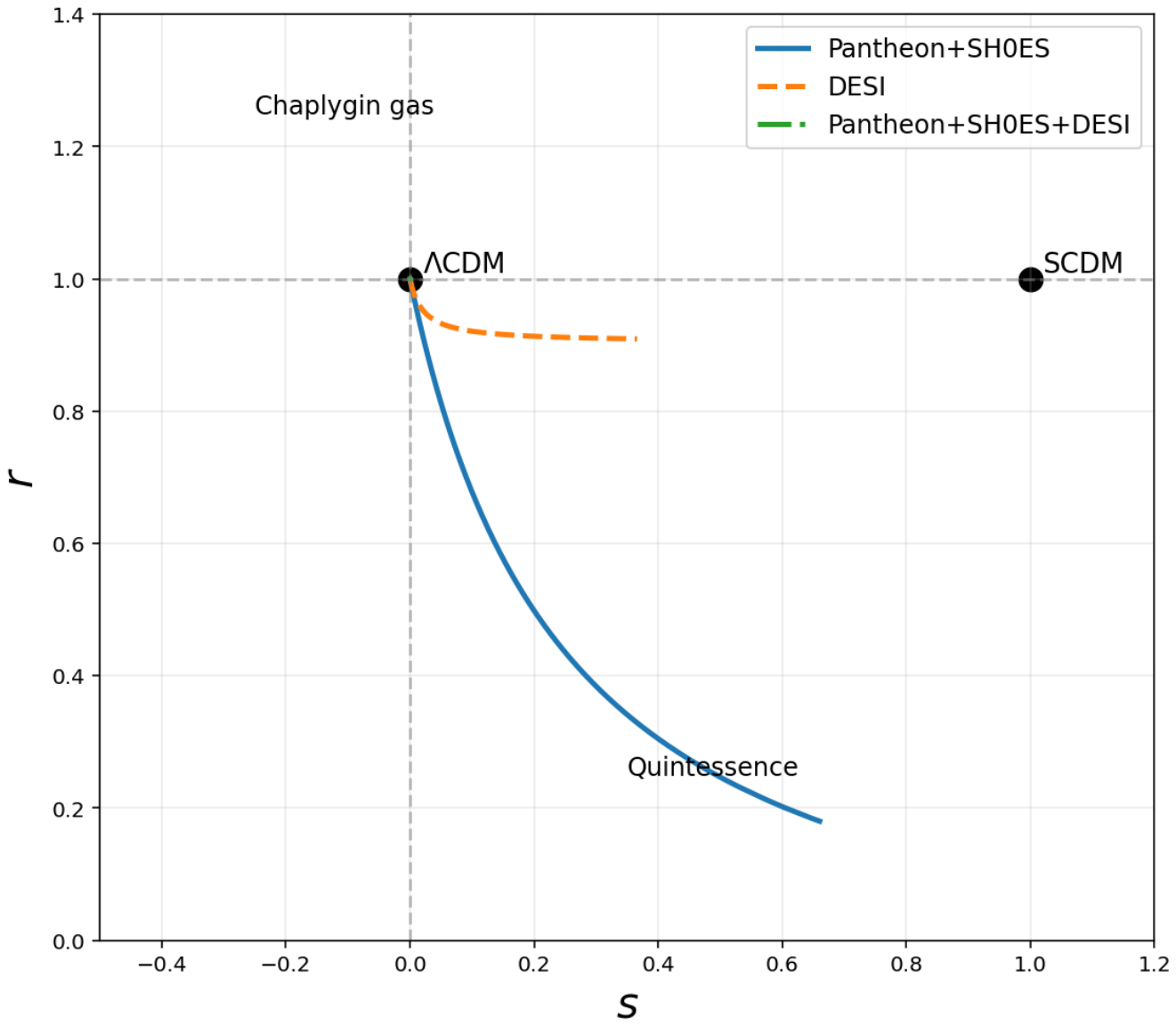}
	\caption{Evolutionary trajectories of the present cosmological model in the $(r,s)$ statefinder plane for the Pantheon+SH0ES, DESI, and Pantheon+SH0ES+DESI datasets. The fixed points corresponding to the $\Lambda$CDM model $(1,0)$ and SCDM $(1,1)$ are also indicated.}
	\label{fig:statefinder}
\end{figure}

To distinguish the present running vacuum model from other cosmological scenarios, we employ the statefinder diagnostic pair $\{r,s\}$ introduced by Sahni et al\cite{Sahni}. The statefinder parameters are constructed from higher-order derivatives of the scale factor and provide an efficient geometrical diagnostic to discriminate among different DE models\cite{Alm}. The statefinder parameter $r(z)$ is defined as
\begin{equation}
	r(z)
	=
	1
	-
	2(1+z)\frac{H'(z)}{H(z)}
	+
	(1+z)^2
	\left[
	\frac{H''(z)}{H(z)}
	+
	\left(
	\frac{H'(z)}{H(z)}
	\right)^2
	\right].
\end{equation}
Using the reconstructed Hubble parameter $H(z)$ substituting into the definition of $r(z)$, we finally obtain the compact form
\begin{equation}
	r(z)
	=
	1
	+
	\frac{
		n(n-3)(H_0^2+A)(1+z)^n
	}
	{
		2H^2(z)
	}.
\end{equation}
The second statefinder parameter is defined as
\begin{equation}
	s(z)
	=
	\frac{r(z)-1}
	{3\left(q(z)-\frac12\right)}.
\end{equation}
Using the analytical form of the deceleration parameter, the corresponding expression for $s(z)$ becomes
\begin{equation}
	s(z)
	=
	\frac{
		n(n-3)(H_0^2+A)(1+z)^n
	}
	{
		3\left[
		(n-3)(H_0^2+A)(1+z)^n
		+
		3A
		\right]
	}.
\end{equation}
The statefinder pair $\{r,s\}$ provides a powerful geometrical diagnostic for distinguishing different DE scenarios beyond the background expansion history. It enables a direct comparison of the present running vacuum model with standard cosmological models such as $\Lambda$CDM, quintessence, Chaplygin gas, and the standard cold dark matter (SCDM) model. In the $\Lambda$CDM cosmology, the statefinder parameters attain the fixed point $(r,s)=(1,0)$, whereas the SCDM model corresponds to $(r,s)=(1,1)$. Consequently, the deviation of the evolutionary trajectories from the $\Lambda$CDM fixed point quantifies the extent to which the present model departs from the standard cosmological scenario. The evolution of the statefinder pair for the best-fit parameters obtained from the Pantheon+SH0ES, DESI, and combined Pantheon+SH0ES+DESI datasets is presented in Fig.~\ref{fig:statefinder}.

It is evident that the trajectory reconstructed from the combined Pantheon+SH0ES+DESI dataset remains closest to the $\Lambda$CDM fixed point throughout its evolution, exhibiting only negligible deviations in both $r$ and $s$. As a result, the trajectory nearly overlaps with the $\Lambda$CDM fixed point and is therefore not visually distinguishable in the main plot. In comparison, the DESI trajectory shows a comparatively larger, although still modest departure. The Pantheon+SH0ES trajectory, however, departs more noticeably from the $\Lambda$CDM fixed point and extends deeper into the quintessence region. This behaviour reflects the comparatively broader parameter constraints obtained from the supernova dataset alone, which permit a wider range of dynamical evolution.

None of the reconstructed trajectories enters the Chaplygin gas region for the considered best-fit parameters. Instead, the present running vacuum model consistently evolves within the quintessence regime while approaching the $\Lambda$CDM fixed point at late times. Overall, the statefinder analysis demonstrates that the proposed model successfully reproduces the geometrical characteristics of the concordance cosmology, with the combined Pantheon+SH0ES+DESI dataset providing the closest agreement with the standard $\Lambda$CDM scenario.
\subsection{$Om$ Diagnostic}
To further distinguish the present cosmological model from the standard $\Lambda$CDM cosmology, we employ the geometrical $Om$ diagnostic introduced in \cite{Sahni2008}. Since the $Om$ diagnostic depends only on the Hubble parameter, it provides a simple and effective probe of the dynamical nature of DE. The $Om(z)$ parameter is defined as
\begin{equation}
	Om(z)=
	\frac{
		\left[\dfrac{H(z)}{H_0}\right]^2-1
	}{
		(1+z)^3-1
	}.
\end{equation}
For the spatially flat $\Lambda$CDM cosmology, $Om(z)$ remains constant and is equal to the present matter density parameter $\Omega_{m0}$. Consequently, any deviation from a constant behaviour indicates a departure from the standard cosmological scenario. Furthermore, the slope of the $Om(z)$ curve provides direct information about the nature of DE: a negative slope corresponds to quintessence-like evolution $(\omega>-1)$, whereas a positive slope indicates phantom behaviour $(\omega<-1)$. Using the analytical expression of the Hubble parameter obtained in the present running vacuum model in $f(R,T)$ gravity, the corresponding $Om(z)$ evolution is reconstructed for the best-fit parameters obtained from the Pantheon+SH0ES, DESI, and the combined Pantheon+SH0ES+DESI datasets.

\begin{figure}
	\centering
	\includegraphics[width=0.75\textwidth]{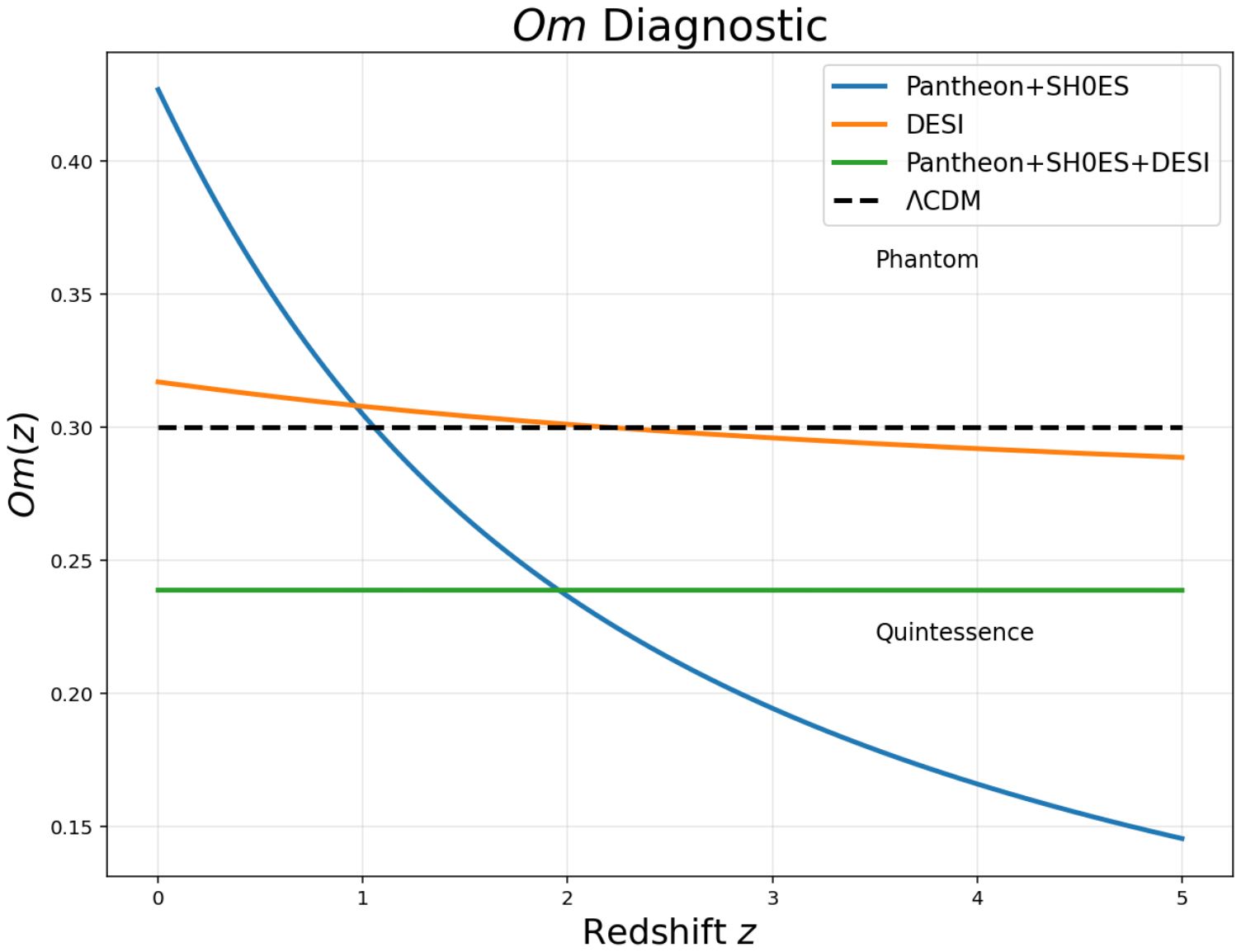}
	\caption{Evolution of the $Om(z)$ diagnostic for the Pantheon+SH0ES, DESI, and Pantheon+SH0ES+DESI datasets. The horizontal dashed line corresponds to the standard $\Lambda$CDM cosmology with $\Omega_{m0}=0.3$. The decreasing nature of the reconstructed trajectories indicates predominantly quintessence-like behavior.}
	\label{fig:om}
\end{figure}

Figure~\ref{fig:om} presents the reconstructed $Om(z)$ trajectories together with the standard $\Lambda$CDM prediction. It is observed that the DESI trajectory remains very close to the $\Lambda$CDM line over the entire redshift range, exhibiting only a slight negative slope. The combined Pantheon+SH0ES+DESI dataset displays an almost constant behaviour with a value slightly lower than the $\Lambda$CDM prediction, indicating that the inclusion of DESI BAO measurements significantly constrains the dynamical evolution of the model. In contrast, the Pantheon+SH0ES trajectory exhibits a noticeably stronger negative slope, crossing the $\Lambda$CDM line near $z\approx1$ and evolving further into the quintessence region at higher redshifts. This comparatively larger deviation reflects the broader parameter constraints obtained from the supernova dataset alone, which allow a wider range of cosmological evolution.

Overall, the reconstructed $Om(z)$ trajectories predominantly exhibit negative or nearly constant slopes throughout the considered redshift range, indicating that the present running vacuum cosmological model evolves mainly within the quintessence regime. The DESI and combined Pantheon+SH0ES+DESI datasets remain in close agreement with the concordance $\Lambda$CDM cosmology, whereas the Pantheon+SH0ES dataset permits comparatively stronger dynamical deviations while remaining observationally consistent with the standard cosmological scenario.
\section{Thermodynamics}
The deep connection between gravity and thermodynamics was first revealed through black hole thermodynamics, where the concepts of black hole entropy and Hawking temperature established a fundamental relationship among gravitation, temperature, and entropy \cite{Beke,Hawking,pad}. A major advancement was made by Jacobson \cite{jac}, who showed that the Einstein field equations can be derived from the fundamental laws of thermodynamics. These developments motivated extensive investigations of thermodynamic laws in the evolution of the universe, where the generalized second law of thermodynamics (GSL) has become an important tool for examining the physical viability of cosmological models and modified theories of gravity.

The GSL requires that the total entropy of the universe, comprising the entropy of the horizon and that of the matter-energy content enclosed within it, must never decrease during the course of cosmic evolution. Since the cosmological dynamics of the present model are formulated in terms of the redshift parameter $z$, it is convenient to express all thermodynamic quantities and their evolution equations within the same framework. Using the transformation $\frac{d}{dt}=-(1+z)H(z)\frac{d}{dz}$, the standard thermodynamic relations expressed in cosmic time are rewritten in terms of the redshift parameter. Following the non-equilibrium thermodynamic formulation in $f(R,T)$ gravity~\cite{momeni}, the total entropy of the universe is expressed as
\begin{equation}
\dot{S}_{total}=\dot{S}_{h}+\dot{S}_{in}+\dot{S}_{prod}
\end{equation}
Here, overdot represents the derivative w.r.t cosmic time. And $\dot{S}_{total}$ denotes the rate of change of the total entropy of the universe, while $\dot{S}_{h}$ and $\dot{S}_{in}$ represent the entropy variation of the apparent horizon and the cosmic fluid enclosed within it, respectively. The term $\dot{S}_{prod}$ corresponds to the additional entropy production arising from the non-equilibrium thermodynamic description of $f(R,T)$ gravity, which arises from the matter--geometry coupling. The generalized second law of thermodynamics is said to be satisfied provided that
$\dot{S}_{total}\geq 0$.

In a spatially flat FLRW universe, the apparent horizon plays a fundamental role in the thermodynamic description of cosmic evolution. The radius of the apparent horizon is given by $R_h=\frac{1}{H}$. Accordingly, the area enclosed by the apparent horizon can be expressed as $A=4\pi R_h^2=\frac{4\pi}{H^2}$.
The entropy associated with the apparent horizon is determined through the entropy-area relation $S_h=\frac{A}{4G_{\rm eff}}$, where $G_{\rm eff}$ represents the effective gravitational coupling. In $f(R,T)$ gravity, the effective gravitational coupling is given by $G_{\rm eff}=\frac{1}{f_R(R,T)}(G+\frac{f_T(R,T)}{8\pi})$. For the specific functional form considered in the present work, $f(R,T)=R+\lambda T$, which yields $f_R=1$ and $f_T=\lambda$. Substituting these relations into the entropy-area expression, the entropy of the apparent horizon becomes $S_h=\frac{8\pi^2}{(\lambda+1)H^2}$. Differentiating the above entropy expression with respect to cosmic time and subsequently expressing it in terms of the redshift parameter yields the evolution equation for the horizon entropy as 
\begin{equation}
\dot S_h  =  \frac{16\pi^2(1+z)}{(\lambda+1) H^2}H'(z),
\end{equation}
where the prime denotes differentiation with respect to the redshift parameter $z$. To determine the entropy variation of the fluid enclosed within the apparent horizon, we employ the Gibbs equation
\begin{equation}
T_{in}\left(dS_{in}+dS_{prod}\right)
=d\left(\rho_{(tot)}V_h\right)+p_{(tot)}dV_h
\end{equation}
where $V_h=\frac{4\pi}{3}R_h^3$ denotes the volume enclosed by the apparent horizon, while $\rho_{tot}$ and $p_{tot}$ represent the effective total energy density and pressure of the cosmic fluid, respectively, including the contributions arising from the matter-geometry coupling. Differentiating the horizon volume with respect to cosmic time and subsequently expressing it in terms of the redshift parameter gives
\begin{equation}
\dot{V}_h=\frac{4\pi(1+z)H'(z)}{H^3(z)}.
\end{equation} Taking the time derivative of Eq.~(49) yields
\begin{equation}
\dot S_{in}+\dot S_{prod} = \frac{(\rho_{(tot)}+p_{(tot)})\dot V_h+V_h\dot\rho_{(tot)}}
	{T_{in}}.
\end{equation}
The temperature of the fluid enclosed by the apparent horizon is assumed to be in thermal equilibrium with the horizon temperature, i.e., $T_{in}=T_h = \frac{2H^2 + \dot H}{4\pi H}$. The temperature associated with the apparent horizon is taken to be the Hayward--Kodama temperature, and can be rewritten as
\begin{equation}
	T_h=\frac{2H(z)-(1+z)H'(z)}{4\pi}.
\end{equation}
which reduces to the Hawking temperature, $T_H=\frac{H}{2\pi}$, in the de-Sitter limit $(\dot H=0)$. The quantities $\rho_{(tot)}$ and $p_{(tot)}$ are appearing in the above equation are obtained from Eqs. (11) and (12), \[
\rho_{(tot)}=\frac{3(1+\lambda)H^{2}-\lambda\dot{H}}
{(1+2\lambda)(1+\lambda)}, \qquad
p_{(tot)}=-\frac{(2+3\lambda)\dot{H}+3(1+\lambda)H^{2}}
{(1+2\lambda)(1+\lambda)}
\] and \[
{\dot\rho_{(tot)}}
=
\frac{6(1+\lambda)H\dot{H}-\lambda\ddot{H}}
{(1+2\lambda)(1+\lambda)}.
\] which can be rewritten in as
\begin{equation}
	\dot{\rho}
	=
	-\frac{
		(1+z)H(z)
		\left[
		(1+z)\lambda\left(H'(z)\right)^2
		+
		H(z)\left(
		(6+7\lambda)H'(z)
		+
		(1+z)\lambda H''(z)
		\right)
		\right]
	}
	{(1+\lambda)(1+2\lambda)}.
\end{equation}
Substituting the transformed relations into Eq.~(50), the entropy variation of the matter-energy content enclosed by the apparent horizon is obtained as
\begin{equation}
	\dot S_{in}+\dot S_{prod}
	=-\frac{
			16\pi^2(1+z)
			\left[
			-(1+z)(6+11\lambda)\left(H'(z)\right)^2
			+
			H(z)\left(
			(6+7\lambda)H'(z)
			+(1+z)\lambda H''(z)
			\right)
			\right]
		}
		{
			3(1+\lambda)(1+2\lambda)\,
			H^2(z)\,
			\left[2H(z)-(1+z)H'(z)\right]
		}.
	\end{equation}
	Using Eq.~(47) and Eq.~(53) into Eq.~(47), the final expression of $\dot{S}_{total}$ obtained as 
	\begin{equation}
	\dot{S}_{total}=	\frac{
			16\pi^2(1+z)
			\left[
			(1+z)(3+5\lambda)\left(H'(z)\right)^2
			+\lambda H(z)
			\left(
			5H'(z)
			-(1+z)H''(z)
			\right)
			\right]
		}
		{
			3(1+\lambda)(1+2\lambda)\,
			H^2(z)\,
			\left[2H(z)-(1+z)H'(z)\right]
		}.
	\end{equation}
	\begin{figure}
		\centering
		\includegraphics[width=0.75\textwidth]{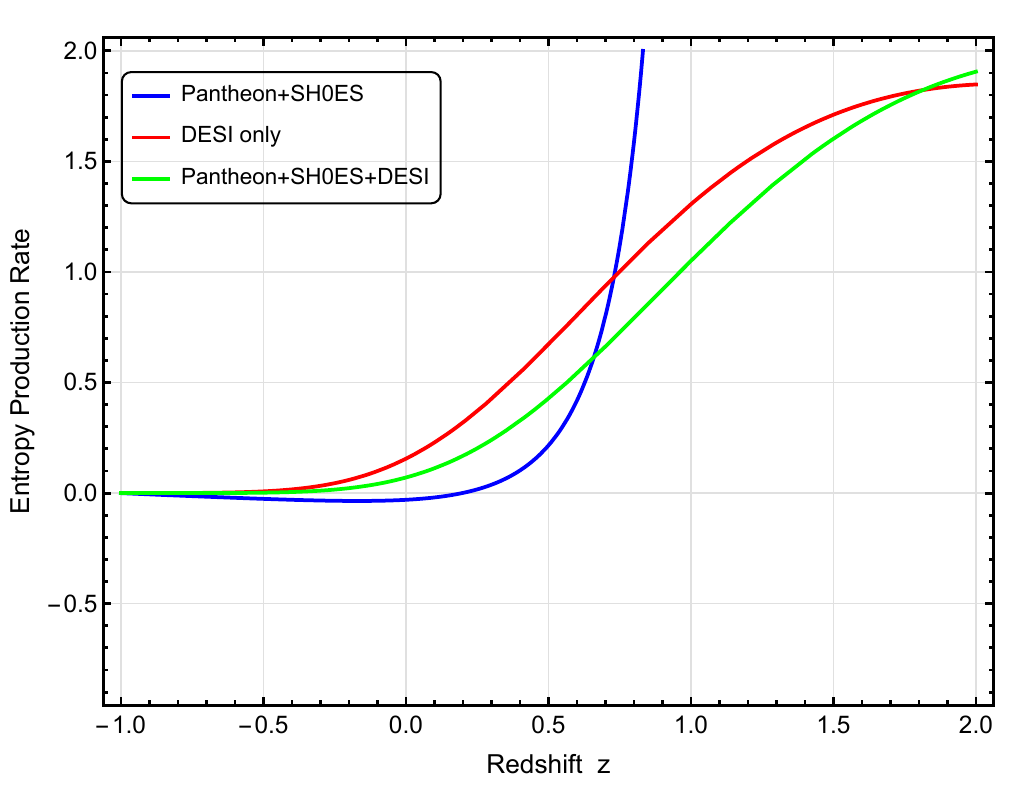}
		\caption{Entropy Production Rate for the Pantheon+SH0ES, DESI, and Pantheon+SH0ES+DESI datasets.}
			\label{fig:Entropy Production Rate}
	
	\end{figure}
 It is evident that the total entropy production rate remains positive throughout almost the entire cosmic evolution for the three observationally constrained datasets, namely Pantheon+SH0ES, DESI, and Pantheon+SH0ES+DESI, as illustrated in Fig.~\ref{fig:Entropy Production Rate}. The entropy production rate reconstructed from the DESI and Pantheon+SH0ES+DESI datasets exhibits a smooth and monotonic evolution without any noticeable irregularities. As the universe evolves toward the far future $(z\rightarrow -1)$, the entropy production rate tends toward a nearly saturated behaviour, indicating that the thermodynamic state approaches asymptotic stability.
 
 A distinct behaviour is observed for the Pantheon+SH0ES dataset, where the entropy production rate exhibits a brief negative excursion around the present epoch $(z\approx0)$. This temporary violation of the GSL is confined to a narrow redshift interval, after which the entropy production rate returns to positive values and continues to increase toward the future. Such behaviour suggests that the departure from thermodynamic equilibrium is only transient and does not affect the overall thermodynamic evolution of the universe. The comparatively larger deviation exhibited by the Pantheon+SH0ES dataset may be attributed to its relatively broader observational constraints obtained from the parameter estimation. Since the entropy production rate explicitly depends on the first and second derivatives of the Hubble parameter, even small variations in the best-fit model parameters can significantly influence these derivative terms, leading to noticeable variations in the entropy production rate. In contrast, the tighter constraints provided by the DESI and the combined Pantheon+SH0ES+DESI datasets result in a smoother thermodynamic evolution throughout the cosmic history.
  
 Overall, the thermodynamic analysis demonstrates that the proposed $f(R,T)$ gravity model with a running vacuum remains largely consistent with the GSL within the observationally constrained parameter space. Although the Pantheon+SH0ES dataset exhibits a temporary violation of the GSL near the present epoch, all three datasets evolve toward a thermodynamically stable future state.
 
\section{Conclusion}

Motivated by the quantum field theoretical description of running vacuum energy, we have investigated a cosmological model in the framework of linear $f(R,T)=R+\lambda T$ gravity, where the vacuum energy density evolves as $\rho_{\Lambda}=\alpha+\beta H^{2}$. In contrast to approaches in which the running vacuum behaviour emerges effectively from modified gravity, the present framework directly incorporates a quantum-inspired running vacuum component while simultaneously accounting for the matter--geometry coupling inherent in $f(R,T)$ gravity. This provides a unified framework for examining the combined influence of vacuum dynamics and modified gravity on the late-time evolution of the universe.

An exact analytical solution for the Hubble parameter was derived from the modified Friedmann equations in a spatially flat FLRW spacetime and constrained using the latest Pantheon+SH0ES Type Ia supernova compilation, DESI BAO measurements, and their combined dataset through the MCMC technique. The resulting parameter constraints indicate that the proposed model remains fully compatible with current late-time cosmological observations. The reconstructed expansion history successfully reproduces the transition from an early decelerating phase to the present accelerated universe, while the effective equation of state remains within the quintessence regime throughout the observed evolution. Furthermore, the reconstructed age of the universe and the residual analysis are both consistent with current observational expectations, providing additional support for the viability of the model.

The geometrical diagnostics further reveal that the proposed running vacuum scenario closely follows the evolutionary behaviour of the concordance $\Lambda$CDM cosmology while preserving its underlying dynamical character. Both the statefinder and Om diagnostics indicate only mild departures from the standard cosmological model. Among the observational datasets considered, the Pantheon+SH0ES reconstruction exhibits comparatively larger deviations, whereas the DESI and, in particular, the combined Pantheon+SH0ES+DESI analysis remain substantially closer to the $\Lambda$CDM behaviour owing to their tighter observational constraints. This demonstrates that the inclusion of complementary late-time observations significantly improves the precision of the model parameters without altering the overall cosmological evolution predicted by the theory.

The thermodynamic analysis further supports the physical consistency of the proposed framework. The GSL is satisfied throughout the cosmic evolution for the DESI and combined Pantheon+SH0ES+DESI best-fit solutions. In contrast, the Pantheon+SH0ES reconstruction exhibits a brief interval around the present epoch during which the total entropy production rate becomes negative, indicating a temporary decrease in the total entropy and, consequently, a local violation of the generalized second law. However, this behaviour is confined to a narrow redshift interval, after which the entropy production rate rapidly becomes positive and approaches a stable asymptotic evolution. This behaviour is fully consistent with the overall observational trend obtained throughout this work, where the comparatively broader constraints from the Pantheon+SH0ES dataset systematically lead to larger deviations in the reconstructed cosmological quantities, whereas the tighter DESI and combined constraints yield a more stable and $\Lambda$CDM-like evolution.

A consistent observational trend emerges throughout the present analysis. The comparatively broader parameter constraints obtained from the Pantheon+SH0ES dataset lead to more pronounced deviations from the concordance $\Lambda$CDM cosmology in the reconstructed expansion history, geometrical diagnostics, and thermodynamic evolution. In contrast, the tighter constraints provided by the DESI observations substantially reduce these deviations, while the combined Pantheon+SH0ES+DESI analysis consistently exhibits the closest agreement with the $\Lambda$CDM scenario. This systematic behaviour demonstrates that the incorporation of complementary late-time cosmological observations not only improves the precision of the model parameters but also enhances the overall robustness and consistency of the proposed cosmological framework.

Overall, the present investigation demonstrates that the combination of a quantum field theory motivated running vacuum with matter--geometry coupling in linear $f(R,T)$ gravity provides a theoretically well-motivated and observationally viable framework for describing the late-time accelerated expansion of the universe. Although the underlying physical mechanism differs fundamentally from that of the standard $\Lambda$CDM cosmology, the resulting cosmological evolution remains remarkably close to the concordance model while preserving the dynamical nature of the vacuum sector and satisfying the essential thermodynamic requirements. The consistency of this behaviour across the reconstructed expansion history, geometrical diagnostics, and thermodynamic analysis demonstrates that the proposed framework provides a coherent description of the late-time universe under the latest observational constraints. Future investigations incorporating additional observational probes, including cosmic microwave background measurements, cosmic chronometers, weak gravitational lensing, and Bayesian model comparison, may further clarify the viability and distinctive predictions of the proposed cosmological scenario.

\section*{Declaration of competing interest} The authors declare that they have no known competing financial interests or personal relationships that could have influenced the work reported in this study.

 \section*{Acknowledgments}
 Author Pankaj Kumar gratefully acknowledges the Inter-University Centre for Astronomy and Astrophysics (IUCAA), Pune, India, for providing research facilities and support under the Visiting Associateship Programme.

\end{document}